\documentclass[aps,pra,amsmath,amssymb,reprint,groupedaddress,showpacs]{revtex4-1}

\usepackage{graphicx}% Include figure files
\usepackage{dcolumn}% Align table columns on decimal point
\usepackage{bm}% bold math

\begin{document}

%\preprint{}

\title{Localized solutions of the Dirac equation in free space and electromagnetic space-time crystals}


\author{G. N. Borzdov}
\email[]{BorzdovG@bsu.by}
%\thanks{}
\affiliation{Department of Theoretical Physics and Astrophysics, Belarusian State University,
4 Nezavisimosti Avenue, 220030 Minsk, Belarus}


%\date{\today}

\begin{abstract}
Localized solutions of the Dirac equation for an electron moving in free space and electromagnetic field lattices with periodic dependence on space-time coordinates (electromagnetic space-time crystals) are treated using the expansions in basis wave functions. The techniques for calculating these functions with any prescribed accuracy are presented. It is shown that in the crystals created by two counterpropagating plane electromagnetic waves with the same or the opposite circular polarizations, the Dirac equation describing the basis functions reduces to matrix ordinary differential equations. These functions and the corresponding mean values of velocity, momentum, energy, and spin operators are found for both types of crystals. Localized solutions describing the families of orthonormal beams in electromagnetic space-time crystals and free space, defined by a given set of orthonormal complex scalar functions on a two-dimensional manifold, are obtained. By way of illustration the orthonormal beams in free space and various localized states with complex vortex structure of probability currents, defined by the spherical harmonics, are presented. The obtained solutions have high probability density only in very small core regions. The evolution of wave packets with one-dimensional localization in both types of crystals created by two circularly polarized waves is described.
\end{abstract}

%\keywords{}

\maketitle

\section{Introduction}
An electromagnetic space-time crystal (ESTC), which is an electromagnetic field lattice with periodic dependence on space-time coordinates, can be created by counterpropagating plane waves. Due to the periodicity of the vector potential in both spatial coordinates and time, one can treat the motion of electrons in ESTCs by analogy with the crystals of solid state physics, described by the Schr\"{o}dinger equation with a periodic electrostatic scalar potential. The natural in this context term ``space-time crystal" was introduced in \cite{gaps}, where the electron wave functions for the ESTC created by two linearly polarized plane waves, were calculated by using the first-order perturbation theory for the Schr\"{o}dinger-Stueckelberg equation. The harmonic time dependence of the Hamiltonian is a generic feature of ESTCs. In a different context, the terms ``time crystal" and ``space-time crystal" have been used  in the recent discussion~\cite{cltime,qtime,ions,watanabe,Yao} around the question of whether time-translation symmetry might be spontaneously broken in a time-independent, conservative classical system and a closed quantum mechanical system, such as chains of trapped ions~\cite{ions,watanabe,Yao}.

In~\cite{bian04,ESTCp1,ESTCp2,ESTCp3,ESTCp4,ESTCp5}, we presented the fundamental solution of the Dirac equation for the ESTC created by six plane waves with the same frequency $\omega_0$ and the four-dimensional wave vectors,
\begin{equation}\label{KNj}
    \bm{K}_\alpha =(k_0 \textbf{e}_\alpha,i k_0), \quad \bm{K}_{\alpha +3}=(-k_0 \textbf{e}_\alpha ,i k_0),
\end{equation}
where $k_0=\omega_0/c=2\pi/\lambda_0$, $c$ is the speed of light in vacuum, $\textbf{e}_\alpha$ are the orthonormal basis vectors, $\alpha =1,2,3$. In this case the periodic vector potential is given by the relation
\begin{equation}\label{field}
  \textbf{A}^\prime \equiv \frac{e}{m_e c^2}\textbf{A}=\sum_{j=1}^6\left(\textbf{A}_j e^{i \bm{K}_j \cdot \bm{x}}+\textbf{A}_j^\ast e^{-i \bm{K}_j \cdot \bm{x}}\right),
\end{equation}
where $e$ is the electron charge, $m_{e}$ is the electron rest mass, $\bm{x}=(\textbf{r},ict)$, $\textbf{r}=x_1\textbf{e}_1+x_2\textbf{e}_2+x_3\textbf{e}_3$, and $x_1$, $x_2$, and $x_3$  are the Cartesian coordinates. The plane waves may have any polarization, so that their complex amplitudes are specified by dimensionless real constants  $a_{jk}$ and $b_{jk}$ as follows:
\begin{equation}\label{abjk}
    \textbf{A}_j = \sum_{k=1}^3 \left(a_{jk} + i b_{jk}\right)\textbf{e}_k, \quad j = 1,2,...,6,
\end{equation}
where $a_{jj} = b_{jj} = a_{j+3\,j} = b_{j+3\,j} = 0, j=1,2,3$. In the general case, Eqs.~(\ref{KNj})--(\ref{abjk}) describe a four-dimensional ESTC (4D-ESTC), i.e., with periodic dependence on all four space-time coordinates. The condition $\textbf{A}_3=\textbf{A}_6=0$ reduces it to a 3D-ESTC with periodic dependence on $x_1, x_2, x_4$, whereas the condition $\textbf{A}_2=\textbf{A}_3=\textbf{A}_5=\textbf{A}_6=0$ results in a 2D-ESTC periodic in $x_1, x_4$. In the simplest case, when $\textbf{A}_1$ is the only nonzero amplitude, the Dirac equation has the well-known Volkov solution~\cite{Volkov}. There exist different representations of this solution~\cite{ESTCp4,Fed,Tern}.

Calculation of quantum electrodynamics (QED) processes occurring in strong laser beams by using the Furry representation \cite{Furry} requires the exact analytical solution of the Dirac equation in the presence of the background electromagnetic field, which describes the so-called laser-dressed electron state. To this end the analytical tools for studying strong-field QED processes in tightly focused laser beams, applying the Wentzel-Kramers-Brillouin (WKB) approximation to find such electron states, have been presented in \cite{Piazza14,Piazza15}. In \cite{Hu2015} the analytical solution for the Klein-Gordon equation in counterpropagating plane waves is obtained, which can be used to derive the solution of the Dirac equation in the same electromagnetic field by the method presented in Ref. \cite{Piazza14}. The techniques presented in \cite{ESTCp1,ESTCp2,ESTCp3,ESTCp4,ESTCp5} provide the tools to find the electron wave functions which describe the laser-dressed states in calculation of QED processes in various ESTCs.

In the last two decades, considerable attention has been focused on the localized solutions of the Dirac equations, such as the free-electron vortex beams carrying orbital angular momentum~\cite{Bliokh2011,Bliokh2017,Birula17,Birula19}. The free-electron vortex states have promising applications in electron microscopy and provide new directions to study fundamental interaction phenomena: (i) the interaction of vortex electrons with intense laser beams and (ii) radiation processes with vortex electrons~\cite{Bliokh2017}. In Ref.~\cite{Bliokh2011} the exact Bessel beam solution of the Dirac equation is constructed from plane wave functions. A different approach to constructing relativistic electron wave packets carrying angular momentum and twisted three-dimensionally localized solutions is suggested in Refs.~\cite{Birula17,Birula19}. In the frame of this approach one starts with a solution of the scalar Klein-Gordon (KG) equation. Then, the bispinor solution of the Dirac equation is expressed in terms of this scalar wave function and its derivatives with spatial coordinates and time.

In Refs. \cite{pre00,*pre01,pre02} we have proposed an approach to designing localized fields, that provides a broad spectrum of tools to construct electromagnetic fields with a high degree of two-dimensional and tree-dimensional spatial localization (2D and 3D localized fields) and promising practical applications. It enables one to obtain a set of orthonormal beams defined by a set of orthonormal scalar functions on a two-dimensional or three-dimensional manifold (beam manifold) and various families of localized fields: three-dimensional standing waves, moving and evolving whirls. In particular, it can be used in designing fields to govern motions of charged and neutral particles. Some illustrations for relativistic electrons in such localized fields have been presented in Ref.~\cite{pre02}. The proposed approach can be applied to any linear field, such as electromagnetic waves in free space, isotropic, anisotropic, and bianisotropic media~\cite{pre00,*pre01,pre02,aeu}, elastic waves in isotropic and anisotropic media~\cite{jpha01a,*jpha01b,*jpha01c,jpha01d}, sound waves~\cite{jpha01d}, weak gravitational waves~\cite{pre01}, etc. In the present article, we extend this approach to the Dirac equation in free space and electromagnetic space-time crystals. In Sec.~\ref{sec:baseq}, we discuss the techniques for calculating the basic wave functions in the ESTCs and free space, mean values of operators velocity, momentum, energy, and spin with respect to these wave functions, and the dispersion relations. Various localized solutions composed of the basis functions are presented in Sec.~\ref{sec:local}.

\section{\label{sec:baseq}Basis functions}
\subsection{4D-ESTC}
\subsubsection{Fundamental solutions}In analysis of localized solutions to the Dirac equation in the ESTCs, there exist two natural units of space-time intervals $\lambda_0=2\pi/k_0$ and $\lambda_e=2\pi/\kappa_e$, related by the parameter
\begin{equation}\label{Om}
   \Omega =\frac{\lambda_e}{\lambda_0}=\frac{k_0}{\kappa_e}=\frac{\hbar \omega_0}{m_e c^2},
\end{equation}
where $\kappa_e=m_{e} c/ \hbar$ and $\hbar$ is the Planck constant. It is convenient to use the dimensionless coordinates $X_k=x_k/\lambda_e, k=1, 2, 3$, $\textbf{R}=\textbf{r}/\lambda_e$, and $X_4=ct/\lambda_e$, so that the Dirac equation takes the form
\begin{eqnarray}\label{diraceq}
  \mathcal{D}\Psi&=&0,\\
  \mathcal{D}&=&\sum_{k=1}^3 \alpha_k\left(-\frac{i}{2\pi}\frac{\partial}{\partial X_k} - A'_k\right) + \alpha_4 - \frac{i}{2\pi} U\frac{\partial}{\partial X_4},\nonumber
\end{eqnarray}
where $U$ is the $4\times 4$ unit matrix, and
\begin{eqnarray}
        \alpha_1=\left(
                            \begin{array}{cccc}
                              0 & 0 & 0 & 1 \\
                              0 & 0 & 1 & 0 \\
                              0 & 1 & 0 & 0 \\
                              1 & 0 & 0 & 0 \\
                            \end{array}
                          \right),\quad
         \alpha_2=\left(
                            \begin{array}{cccc}
                              0 & 0 & 0 & -i \\
                              0 & 0 & i & 0 \\
                              0 & -i & 0 & 0 \\
                              i & 0 & 0 & 0 \\
                            \end{array}
                          \right),\nonumber\\
        \alpha_3=\left(
                            \begin{array}{cccc}
                              0 & 0 & 1 & 0 \\
                              0 & 0 & 0 & -1 \\
                              1 & 0 & 0 & 0 \\
                              0 & -1 & 0 & 0 \\
                            \end{array}
                          \right),\quad
        \alpha_4=\left(
                            \begin{array}{cccc}
                              1 & 0 & 0 & 0 \\
                              0 & 1 & 0 & 0 \\
                              0 & 0 & -1 & 0 \\
                              0 & 0 & 0 & -1 \\
                            \end{array}
                          \right)\nonumber.
\end{eqnarray}

For a given four-dimensional wave vector,
\begin{equation}\label{Kkq}
    \bm{K}=(\textbf{k},i\omega/c)=\kappa_e \textbf{Q},\quad \textbf{Q},=(\textbf{q},i q_4),
\end{equation}
Eq.~(\ref{diraceq}) has the solution~\cite{ESTCp1,ESTCp4,ESTCp5}:
\begin{equation}\label{sol1}
    \Psi=\Psi_0 e^{i \bm{x}\cdot\bm{K}},\quad \Psi_0=\sum_{n \in \mathcal L} c(n) e^{i \bm{x}\cdot\bm{G}(n)},
\end{equation}
where
\begin{equation}\label{qq4}
\textbf{q} = q_1\textbf{e}_1+q_2\textbf{e}_2+q_3\textbf{e}_3 = \frac{\hbar \textbf{ k}}{m_e c}, \quad   q_4 = \frac{\hbar \omega}{m_e c^2},
\end{equation}
\begin{eqnarray}\label{xKxGn}
  \bm{x}\cdot\bm{K}&=&2\pi(\textbf{q}\cdot\textbf{R}-q_4 X_4),\nonumber\\
  \bm{x}\cdot\bm{G}(n)&=&2\pi\Omega(\textbf{n}\cdot\textbf{R}-n_4 X_4),
\end{eqnarray}
$\bm{G}(n) = (k_0 \textbf{n},i k_0 n_4)$, $\textbf{n} = n_1\textbf{e}_1+n_2\textbf{e}_2+n_3\textbf{e}_3$, points $n=(n_1,n_2,n_3,n_4)$ of the integer lattice $\mathcal L$ have even values of the sum $n_1+n_2+n_3+n_4$, and $c(n)$ are the Fourier amplitudes (bispinors). The bispinor function $\Psi_0$ is periodic in $X_1, X_2, X_3$, and $X_4$ with the period $\tau=1/\Omega$.

Substitution of $\textbf{A}^\prime$ (\ref{field}) and $\Psi$ (\ref{sol1}) in Eq.~(\ref{diraceq}) results in the infinite system of homogeneous matrix equations relating bispinors $c(n)$~\cite{ESTCp1,ESTCp4,ESTCp5}. In the general case, each amplitude $c(n)$ enters in 13 different equations of this system. The set $C =\{c(n),n \in {\mathcal L}\}$ of the Fourier amplitudes $c(n)$ can be treated as an element of an infinite-dimensional complex linear space $V_C$. Since for any given $n \in {\mathcal L}$, $c(n)$ is the bispinor, $C \in V_C$ is called the multispinor. The fundamental solution  $\mathcal{S}$ is the Hermitian operator of projection ($\mathcal{S}^\dag=\mathcal{S}^2=\mathcal{S}$) onto the solution subspace of the multispinor space $V_C$. For any $C_0\in V_C$, $C=\mathcal{S}C_0$ is the exact particular solution specified by the multispinor $C_0$, i.e., the function $\Psi$ Eq.~(\ref{sol1}) with the set of amplitudes $\{c(n), n \in \mathcal{L}\}=\mathcal{S}C_0$  satisfies Eq.~(\ref{diraceq}) for the problem under consideration. The fundamental solution $\mathcal{S}$ has been expressed in terms of an infinite series of projection operators calculated by a recurrent process based on a fractal approach. This technique has been detailed and applied to various ESTCs in Refs.~\cite{ESTCp1,ESTCp2,ESTCp3,ESTCp4,ESTCp5}.

\subsubsection{Particular solutions}
Numerical implementation of the presented technique implies the replacement of the infinite system of matrix equations by its subsystem defined by some finite solution domain ${\mathcal L'} \subset \mathcal{L}$, whereas outside $\mathcal L'$ bispinor amplitudes $c(n)$ are assumed vanishing. The  recurrent process gives the exact fundamental solution of the subsystem, which is an approximate solution of the corresponding infinite system. In this case, the electron wave function is approximated by a bispinor function with a limited discrete Fourier spectrum. When the amplitude $C_0$ is localized at the point $n_o=(0,0,0,0)$, it is specified by one bispinor as $C_0=\{c(n_o)\}$ and the solution domain can be given as
\begin{equation}\label{Lgmax}
    \mathcal{L'}=\{n=(n_1,n_2,n_3,n_4), 0\leq g_{4d}(n)\leq g_{\rm max}\},
\end{equation}
where $g_{4d}(n_1,n_2,n_3,n_4)=\max\{|n_1|+|n_2|+|n_3|,|n_4|\}$ and $g_{\rm max}$ is the integer specifying the domain size and hence the accuracy of such approximations~\cite{ESTCp3,ESTCp4,ESTCp5}.

For an approximate solution
\begin{equation}\label{sol1apr}
    \Psi=\Psi'_0 e^{i \bm{x}\cdot\bm{K}},\quad \Psi'_0=\sum_{n \in \mathcal L'} c'(n) e^{i \bm{x}\cdot\bm{G}(n)},
\end{equation}
the functional
\begin{equation}\label{Rpsi}
    R: \Psi\mapsto R[\Psi]=\frac{\|\mathcal{D}\Psi\|}{\|\Psi\|},
\end{equation}
where
\begin{equation}\label{normpsi}
    \|\Psi\|=\left(\sum_{n \in \mathcal L'}c'^{\dag}(n)c'(n)\right)^{1/2},
\end{equation}
evaluates the relative residual at the substitution of $\Psi$ Eq. (\ref{sol1apr}) into Eq.~(\ref{diraceq}). It provides a convenient fitness criterion to  accurately compare various approximate solutions of this equation~\cite{ESTCp2,ESTCp3,ESTCp4,ESTCp5}. If $R[\Psi]\ll 1$, then the function $\Psi$ may be treated as a reasonable approximation to the exact solution for which $R[\Psi]=0$, and the smaller is $R[\Psi]$, the more accurate is the approximation.

Since the system of matrix equations in $c(n)$ is homogeneous, $\textbf{q}$ and $q_4$ are related by a dispersion relation which manifests itself in the spectral distribution of $c(n)$ for each exact particular solution $\Psi$~(\ref{sol1}). In the general case, this Fourier spectrum is nonlocalized. In numerical calculations instead of an exact particular solution, we obtain its approximation with a localized Fourier spectrum bounded by the truncation condition $ g_{4d}(n)\leq g_{\rm max}$ for all $n\in\mathcal{L}'$. Consequently, the dispersion interrelation of $\textbf{q}$ and $q_4$ is defined by the minimum of the fitness function $R(\xi)=R[\Psi(\bm{x},\xi)]$ with graphical representation in the form of a spectral curve of approximate solutions~\cite{ESTCp3,ESTCp4,ESTCp5}, where
\begin{equation}\label{xiq4}
    \xi=q_4-\sqrt{1+\textbf{q}^2}=\frac{\hbar \omega}{m_e c^2}-\sqrt{1+\left(\frac{\hbar \textbf{k}}{m_e c}\right)^2}.
\end{equation}
The ESTCs created by circularly polarized waves possess the spin birefringence, i.e., at a given quasimomentum $\textbf{q}$, the dispersion equation has two solutions $q_{4a}$ and $q_{4b}$,  which specify bispinor wave functions describing electron states with different energies and mean values of momentum and spin operators~\cite{ESTCp3,ESTCp4,ESTCp5}.

\subsubsection{\label{sec:orthorel}Orthogonality relation}
Let us consider two solutions of Eq.~(\ref{diraceq})
\begin{equation}\label{solab}
    \Psi_a=\Psi_{0a} e^{i \bm{x}\cdot\bm{K}_a},\quad \Psi_b=\Psi_{0b} e^{i \bm{x}\cdot\bm{K}_b},
\end{equation}
where $\bm{K}_a=(\kappa_e \textbf{q}_a,i\kappa_e q_{4a})$ and $\bm{K}_b=(\kappa_e \textbf{q}_b,i\kappa_e q_{4b})$. Substitution of $\Psi_a$ in Eq.~(\ref{diraceq}) results in the identity $\mathcal{D}_{0a}\Psi_{0a}\equiv 0$, where
\begin{eqnarray}\label{mathD0a}
  \mathcal{D}_{0a}&=&\sum_{k=1}^3 \alpha_k\left[-\frac{i}{2\pi}\frac{\partial}{\partial X_k} +(q_{ka} - A'_k)U\right]\nonumber\\
  &+&\alpha_4 -  q_{4a}U- \frac{i}{2\pi} U\frac{\partial}{\partial X_4}.
\end{eqnarray}
In a similar manner, $\mathcal{D}_{0b}\Psi_{0b}\equiv 0$ and hence the identity $(\Psi_{0a}^{\dag}\mathcal{D}_{0b}\Psi_{0b})^*-\Psi_{0b}^{\dag}\mathcal{D}_{0a}\Psi_{0a}\equiv 0$ takes the form
\begin{eqnarray}\label{idini}
  \sum_{k=1}^3 \frac{i}{2\pi}\frac{\partial}{\partial X_k}(\Psi_{0b}^{\dag}\alpha_k\Psi_{0a}) &+&  \sum_{k=1}^3  (q_{kb} - q_{ka} )\Psi_{0b}^{\dag}\alpha_k\Psi_{0a}\nonumber\\
   -  (q_{4b} - q_{4a})\Psi_{0b}^{\dag}\Psi_{0a} &+& \frac{i}{2\pi} \frac{\partial}{\partial X_4}(\Psi_{0b}^{\dag}\Psi_{0a})\equiv 0.
\end{eqnarray}
Let now $\bm{K}_a$ and $\bm{K}_b$ be two different solutions of the dispersion equation and
\begin{equation}\label{Psi0ab}
    \Psi_{0a}=\sum_{n \in \mathcal L} a(n) e^{i \bm{x}\cdot\bm{G}(n)}, \Psi_{0b}=\sum_{m \in \mathcal L} b(m) e^{i \bm{x}\cdot\bm{G}(m)}.
\end{equation}
Taking into account the periodicity of these amplitude functions, upon integrating Eq.~(\ref{idini}) over $X_1, X_2, X_3$, and $X_4$ from $0$ to $\tau$, we obtain the orthogonality relation
\begin{equation}\label{orthorel}
    \sum_{n \in \mathcal L} b^{\dag}(n)\left[\sum_{k=1}^3(q_{kb} - q_{ka})\alpha_k - (q_{4b} - q_{4a})U\right]a(n) = 0.
\end{equation}
If $\textbf{q}_a=\textbf{q}_b=\textbf{q}$ and $q_{4b}\neq q_{4a}$, then it reduces to~\cite{ESTCp4}
\begin{equation}\label{orthoq}
    \sum_{n \in \mathcal L} b^{\dag}(n)a(n) = 0.
\end{equation}

\subsection{\label{sec:2D}2D-ESTC}
\subsubsection{Evolution equations}
In the 2D-ESTC, the potential $\textbf{A}^\prime$ Eq.~(\ref{field}) takes the form:
\begin{eqnarray}\label{field2D}
  \textbf{A}^\prime&=&\textbf{A}_1 e^{i(\varphi_1-\varphi_4)}+\textbf{A}_1^\ast e^{i(-\varphi_1+\varphi_4)}\nonumber\\
 &+&\textbf{A}_4 e^{i(-\varphi_1-\varphi_4)}+\textbf{A}_4^\ast e^{i(\varphi_1+\varphi_4)},
\end{eqnarray}
where $\varphi_j=2\pi\Omega X_j, j=1,2,3,4$. For this case, we present below two families of solutions to the Dirac equation,
\begin{equation}\label{VWX4}
    \Psi=\Psi_0 e^{2\pi i q_1X_1}, \Psi_0=V(X_4)+i e^{i s\varphi_1}W(X_4),
\end{equation}
and
\begin{equation}\label{VWX1}
    \Psi=\Psi_0 e^{-2\pi i q_4X_4}, \Psi_0=V(X_1)+i e^{-i s\varphi_4}W(X_1),
\end{equation}
where $s=\pm 1$. Substitutions of Eqs.~(\ref{VWX4}) and (\ref{VWX1}) in Eq.~(\ref{diraceq}) result in two ordinary differential equations which have nonzero solutions only in the 2D-ESTCs created by circularly polarized waves ($\textbf{A}_1^2=\textbf{A}_4^2=0$). Without the loss of generality, the appropriate amplitudes can be written as follows:
\begin{equation}\label{A1A4}
    \textbf{A}_1=a_{12}(\textbf{e}_2+ig\textbf{e}_3),\quad \textbf{A}_4=a_{42}(\textbf{e}_2+igL\textbf{e}_3),
\end{equation}
where $a_{12}$ and $a_{42}$ are real numbers, $g=\pm 1$. The solutions $\Psi$ Eq.~(\ref{VWX4}) exist when the counterpropagating waves have the same circular polarization ($L=-1$), whereas the solutions $\Psi$ Eq.~(\ref{VWX1}) exist in the case of the waves with left and right circular polarizations    ($L=1$).

The bispinor functions $V=V(X_4)$ and $W=W(X_4)$ satisfy the evolution equation
\begin{equation}\label{dVWdX4}
    \frac{\rm d}{{\rm d}X_4}\left(\begin{array}{c}
                              V \\
                              W
                            \end{array}\right)
    =2\pi i \left(\begin{array}{cc}
             a & b \\
             c & d
           \end{array}\right)\left(\begin{array}{c}
                        V \\
                        W
                      \end{array}\right),
\end{equation}
where
\begin{eqnarray}\label{abcd4}
  a&=&-q_1\alpha_1-\alpha_4,\quad d=-(q_1+s\Omega)\alpha_1-\alpha_4,\nonumber\\
  b&=&c^\dag=(i\alpha_2+p\alpha_3)\left(a_{12}e^{is\varphi_4}+a_{42}e^{-is\varphi_4}\right),
\end{eqnarray}
and $p=gs$. The functions $V=V(X_1)$ and $W=W(X_1)$ satisfy the equation
\begin{equation}\label{dVWdX1}
    \left(\begin{array}{cc}
             \alpha_1 & 0 \\
             0 & \alpha_1
           \end{array}\right)\frac{\rm d}{{\rm d}X_1}\left(\begin{array}{c}
                              V \\
                              W
                            \end{array}\right)
    =2\pi i \left(\begin{array}{cc}
             a & b \\
             c & d
           \end{array}\right)\left(\begin{array}{c}
                        V \\
                        W
                      \end{array}\right),
\end{equation}
where
\begin{eqnarray}\label{abcd1}
  a&=&q_4 U-\alpha_4,\quad d=(q_4+s\Omega)U-\alpha_4,\nonumber\\
  b&=&c^\dag=(i\alpha_2+p\alpha_3)\left(a_{12}e^{-is\varphi_1}+a_{42}e^{is\varphi_1}\right).
\end{eqnarray}

Both the families of solutions are subject to the same conditions,
\begin{equation}\label{VWspace}
    (i\alpha_2+p\alpha_3)V=0,\quad (i\alpha_2-p\alpha_3)W=0,
\end{equation}
which can be taken properly into account using the basis
\begin{eqnarray}\label{u1234}
  u_1&=&\frac{1}{\sqrt{2}}\left(\begin{array}{c}
          1 \\
          1 \\
          0 \\
          0
        \end{array}\right),\quad u_2=\frac{1}{\sqrt{2}}\left(\begin{array}{c}
          1 \\
          -1 \\
          0 \\
          0
        \end{array}\right),\nonumber\\
  u_3&=&\frac{1}{\sqrt{2}}\left(\begin{array}{c}
          0 \\
          0 \\
          1 \\
          1
        \end{array}\right),\quad u_4=\frac{1}{\sqrt{2}}\left(\begin{array}{c}
          0 \\
          0 \\
          1 \\
          -1
        \end{array}\right).
\end{eqnarray}
Since
\begin{equation}\label{u2341}
    \frac{1}{2}(\alpha_3+i\alpha_2)=\frac{1}{2}(\alpha_3-i\alpha_2)^\dag=u_2\otimes u_3^\dag + u_4\otimes u_1^\dag,
\end{equation}
the bispinors $V$ and $W$ can be written as
\begin{equation}\label{VWpminus1}
    V=V_1u_1+V_3u_3,\quad  W=W_2u_2+W_4u_4
\end{equation}
for $p=-1$, and
\begin{equation}\label{VWpplus1}
    V=V_2u_2+V_4u_4,\quad  W=W_1u_1+W_3u_3
\end{equation}
for $p=1$. Because of this, Eqs.~(\ref{dVWdX4}) and (\ref{dVWdX1}) reduce to the similar evolution equations,
\begin{equation}\label{dZdXj}
    \frac{\rm d}{{\rm d}X_j}Z
    =2\pi i M_j Z,
\end{equation}
where $j=1$ for $L=1$, and $j=4$ for $L=-1$,
\begin{equation}\label{ZZ}
    Z=\left(\begin{array}{c}
          V_1 \\
          V_3 \\
          W_2 \\
          W_4
        \end{array}\right)\text{for } p=-1,  Z=\left(\begin{array}{c}
          V_2 \\
          V_4 \\
          W_1 \\
          W_3
        \end{array}\right)\text{for } p=1,
\end{equation}
\begin{equation}\label{M1}
    M_1=\left(
          \begin{array}{cccc}
            0 & -pq_4-p & -F_{1,s}^* & 0 \\
            -pq_4+p & 0 & 0 & -F_{1,s}^* \\
            F_{1,s} & 0 & 0 & pq'_4 +p \\
            0 & F_{1,s} & pq'_4 -p & 0 \\
          \end{array}
        \right),
\end{equation}
\begin{equation}\label{M4}
    M_4=\left(
          \begin{array}{cccc}
            -1 & pq_1 & 0 & pF_{4,s} \\
            pq_1 & 1 & pF_{4,s} & 0 \\
            0 & pF_{4,s}^* & -1 & -pq'_1 \\
            pF_{4,s}^* & 0 & -pq'_1 & 1 \\
          \end{array}
        \right),
\end{equation}
\begin{equation}\label{Fjs}
    F_{j,s}=2\left(a_{12}e^{is\varphi_j}+a_{42}e^{-is\varphi_j}\right),
\end{equation}
$q'_j=q_j+s\Omega$ and $F_{j,s}^*\equiv F_{j,-s}$.

\subsubsection{Fundamental solutions}
The evolution operator ${\mathcal F}_j$ [the fundamental solution of Eq.~(\ref{dZdXj})] describes the dependence $Z$ on $X_j$ for the whole family of particular solutions as
\begin{equation}\label{ZFZ0}
    Z={\mathcal F}_j(X_j)Z(0),
\end{equation}
where $Z(0)$ may be prescribed arbitrarily and
\begin{equation}\label{FjXj}
    {\mathcal F}_j(X_j)=\stackrel{\Leftarrow}{\int\nolimits_0^{X_j}}(U+2\pi i M_j dX_j)
\end{equation}
is a multiplicative integral. The multiplicative integral of a matrix function $P=P(t)$ is defined as follows~\cite{Gant}:
\begin{eqnarray}  \label{UPdt}
\stackrel{\Leftarrow}{\int\nolimits_{t_0}^{t}}(U+Pdt)&& \nonumber\\
=\lim_{{\Delta}t_k{\rightarrow}0}
&&\left[U+P(t_n)\Delta{t_n}\right]\ldots\left[U+P(t_1)\Delta{t_1}\right],\hspace{10pt}
\end{eqnarray}
where $U$ is the unit matrix, $t_1, t_2, \ldots, t_{n-1}$ are arbitrary intermediate points splitting the interval $[t_0,t]$ into $n$ parts, ${\Delta}t_k=t_k-t_{k-1}$, $k=1,2,\ldots,n; t_n=t$. If the matrix $P$ is independent of $t$, then this integral reduces to the exponential $\exp[(t-t_0)P]$. The above definition provides the direct way to close numerical approximations of multiplicative integrals by using sufficiently small steps ${\Delta}t_k$. However, in analytical investigation and numerical calculation of evolution operators for superpositions of counterpropagating waves, like electromagnetic fields in a plane stratified bianisotropic medium, it is useful first to apply the wave splitting technique~\cite{jmp97} based on the integration by parts for multiplicative integrals~\cite{Gant}. In particular, it reduces multiplicative integrals of $4\times 4$ matrix functions with strong (``fast") dependence on the integration variable to multiplicative integrals of $2\times 2$ matrix functions with weak (``slow") dependence on this variable.

\subsubsection{Particular solutions}
Owing to the periodicity of the matrix functions $M_1=M_1(X_1)$ Eq. (\ref{M1}) and $M_4=M(X_4)$ Eq. (\ref{M4}), from the Lyapunov theorem~\cite{Gant} it follows that ${\mathcal F}_j$ Eq.~(\ref{FjXj}) can be written as
\begin{equation}\label{FjPjCj}
    {\mathcal F}_j={\mathcal P}_j(X_j)e^{2\pi i X_j C_j},
\end{equation}
where ${\mathcal P}_j$ is a periodic matrix function in $X_j$ with the period $\tau$, $C_j$ is a matrix independent of $X_j$, and $j=1, 4$. It is significant that $M_j$ and hence $C_j$ are specified by a given $q_n$, where $n=5-j$. We use below these related indices to describe both the families of solutions in a concise form.

Let us consider the particular solutions of Eq.~(\ref{dZdXj}) defined by the eigensystem of matrix $C_j$ as $C_j Z(0)=\eta_j Z(0)$, $\eta_1\equiv q_1$ and $\eta_4\equiv -q_4$. In this case, from Eqs.~(\ref{ZFZ0}) and (\ref{FjPjCj}) follows
\begin{equation}\label{ZYXj}
    Z=Y e^{2\pi i \eta_j X_j},
\end{equation}
where $Y={\mathcal P}_j(X_j) Z(0)$ is the periodic function in $X_j$ with the period $\tau$ and hence it can be written as
\begin{equation}\label{YYk}
    Y=\left(\begin{array}{c}
          y_1 \\
          y_2 \\
          y_3 \\
          y_4
        \end{array}\right)=\sum_{k=-\infty}^{+\infty}Y_k e^{ik\varphi_j}.
\end{equation}
Substituting these relations in Eqs.~(\ref{VWX4}), (\ref{VWX1}), and (\ref{VWpminus1})--(\ref{ZZ}) gives:
\begin{eqnarray}\label{psiq1q4}
  \Psi=e^{2\pi i(q_1X_1-q_4X_4)} \left\{V_0(\varphi_j)\right. \nonumber  \\
   \left.+i W_0(\varphi_j)\exp\left[(-1)^j i s \varphi_n\right]\right\},
\end{eqnarray}
where
\begin{equation}\label{V0W0}
    V_0=y_1u_{\alpha}+y_2u_{\beta},\quad W_0=y_3u_{\gamma}+y_4u_{\delta},
\end{equation}
\begin{eqnarray}\label{abcdforp}
  \{\alpha,\beta,\gamma,\delta\}&=&\{1,3,2,4\} \text{ for } p=-1,\nonumber\\
  &=&\{2,4,1,3\} \text{ for } p=1.
\end{eqnarray}
In the 2D-ESTC created by two waves with the same circular polarization [$L=-1$ in Eq.~(\ref{A1A4}), $j=4$, and $n=1$] $M_4, C_4$, and $q_4$ depend on $q_1$, hence, this defines a dispersion relation $q_4=q_4(q_1)$. In the 2D-ESTC created by two waves with left and right circular polarizations ($L=1$, $j=1$, and $n=4$) $M_1, C_1$, and $q_1$ depend on $q_4$, and this defines a different dispersion relation $q_1=q_1(q_4)$.

The matrix functions $M_1$ and $M_4$ can be written as
\begin{equation}\label{Mj0pm}
    M_j=M_{j0}+M_{j-}e^{-i\varphi_j}+M_{j+}e^{i\varphi_j},
\end{equation}
where $j=1, 4$, and
\begin{equation}\label{M10}
    M_{10}=p\left(
          \begin{array}{cccc}
            0 & -q_4-1 & 0 & 0 \\
            -q_4+1 & 0 & 0 & 0 \\
            0 & 0 & 0 & q'_4 +1 \\
            0 & 0 & q'_4 -1 & 0 \\
          \end{array}
        \right),
\end{equation}
\begin{equation}\label{M1pm}
    M_{1\pm}=p\left(
          \begin{array}{cccc}
            0 & 0 & -f_{\mp} & 0 \\
            0 & 0 & 0 & -f_{\mp} \\
            f_{\pm} & 0 & 0 & 0 \\
            0 & f_{\pm} & 0 & 0 \\
          \end{array}
        \right),
\end{equation}
\begin{equation}\label{M40}
    M_{40}=\left(
          \begin{array}{cccc}
            -1 & pq_1 & 0 & 0 \\
            pq_1 & 1 & 0 & 0 \\
            0 & 0 & -1 & -pq'_1 \\
            0 & 0 & -pq'_1 & 1 \\
          \end{array}
        \right),
\end{equation}
\begin{equation}\label{M4pm}
    M_{4\pm}=\left(
          \begin{array}{cccc}
            0 & 0 & 0 & f_{\pm} \\
            0 & 0 & f_{\pm} & 0 \\
            0 & f_{\mp} & 0 & 0 \\
            f_{\mp} & 0 & 0 & 0 \\
          \end{array}
        \right),
\end{equation}
\begin{equation}\label{fpm}
    f_{\pm}=g[a_{12}(s \pm 1)+a_{42}(s \mp 1)].
\end{equation}
Substitution of $Z$ Eq.~(\ref{ZYXj}) in Eq.~(\ref{dZdXj}) results in the infinite system of matrix equations
\begin{equation}\label{djkeq}
    Q_j(k)=0,\quad k=0, {\pm 1}, {\pm 2},\ldots,
\end{equation}
where
\begin{eqnarray}\label{djkNjk}
  Q_j(k)&=&N_j(k)Y_k+M_{j+}Y_{k-1}+M_{j-}Y_{k+1},\nonumber \\
  N_j(k)&=&M_{j0}-(\eta_j+k\Omega)U.
\end{eqnarray}
It can also be written as
\begin{equation}\label{YkTY}
    Y_k=T_{-}(k)Y_{-1}+T_{+}(k)Y_{1},\quad k=0, {\pm 1}, {\pm 2},\ldots,
\end{equation}
where $T_{-}(-1)=T_{+}(1)=U$, $T_{-}(1)=T_{+}(-1)=0$, and
\begin{equation}\label{T0pm}
    T_{\mp}(0)=-[N_j(0)]^{-1}M_{j\pm},
\end{equation}
\begin{equation}\label{Tkm}
    T_{\mp}(k-1)=-M_{j+}^{-1}[N_j(k)T_{\mp}(k)+M_{j-}T_{\mp}(k+1)],
\end{equation}
for $k=-1, -2,\ldots$, and
\begin{equation}\label{Tkp}
    T_{\mp}(k+1)=-M_{j-}^{-1}[N_j(k)T_{\mp}(k)+M_{j+}T_{\mp}(k-1)],
\end{equation}
for $k=1, 2,\ldots$. Because of this, the particular solution $Z$ Eq.~(\ref{ZYXj}) and hence the wave function $\Psi$ Eq.~(\ref{psiq1q4}) are uniquely defined by $\eta_j$ and the Fourier amplitudes $Y_{\mp 1}$.

To find these parameters, let us replace the exact solution $Z$ by an approximate solution
\begin{equation}\label{Zprime}
    Z'=\sum_{k=-k_m}^{k_m}Y_k e^{2\pi i(\eta_j+k\Omega)X_j},
\end{equation}
obtained from Eqs.~(\ref{ZYXj}) and (\ref{YYk}) by the truncation conditions $Y_k=0$ and $T_{\mp}(k)=0$ for $|k|>k_m$. For this function, Eqs.~(\ref{djkNjk})--(\ref{Tkp}) result in the identities $Q_j(k)\equiv 0$ for $|k|\leq k_m -1$ and $|k|> k_m +1$, whereas
\begin{eqnarray}\label{Qres}
  Q_j(\pm k_m)&=&N_j(\pm k_m)Y_{\pm k_m}+M_{j\pm}Y_{\pm(k_m -1)}, \nonumber\\
  Q_j(k_m +1)&=&M_{j+}Y_{k_m}, \nonumber\\
  Q_j(-k_m -1)&=&M_{j-}Y_{-k_m}
\end{eqnarray}
remain nonzero. The norm of $Z'$ can be written as
\begin{equation}\label{Zpnorm}
    \|Z'\|=\left({\mathcal Y}^{\dag}{\mathcal N}{\mathcal Y}\right)^{1/2},\quad  {\mathcal Y}=\left(\begin{array}{c}
          Y_{-1} \\
          Y_1
        \end{array}\right),
\end{equation}
where
\begin{equation}\label{Ncal}
    {\mathcal N}=\sum_{k=-k_m}^{k_m}\left(\begin{array}{c}
                                            T_{-}^{\dag}(k) \\
                                            T_{+}^{\dag}(k) \\
                                            \end{array}
                                            \right)\left(
                                            \begin{array}{cc}
                                            T_{-}(k) & T_{+}(k) \\
                                            \end{array}
                                            \right).
\end{equation}
Since Eq.~(\ref{dZdXj}) can be rewritten as $d_j Z=0$, where
\begin{equation}\label{djMj}
    d_j=M_j+\frac{i}{2\pi}U\frac{\rm d}{{\rm d}X_j},
\end{equation}
the relative residual $R$ at the substitution of $Z'$ into this equation is defined by the relation
\begin{equation}\label{RZprime}
    R=\frac{\|d_j Z'\|}{\|Z'\|}=\left(\frac{{\mathcal Y}^{\dag}{\mathcal R}{\mathcal Y}}{{\mathcal Y}^{\dag}{\mathcal N}{\mathcal Y}}\right)^{1/2},
\end{equation}
where
\begin{equation}\label{Rcal}
    {\mathcal R}={\mathcal S}(-k_m -1)+{\mathcal S}(-k_m)+{\mathcal S}(k_m)+{\mathcal S}(k_m+1),
\end{equation}
\begin{equation}\label{Scal}
    {\mathcal S}(k)=\left(
                   \begin{array}{c}
                     D_{-}^{\dag}(k) \\
                     D_{+}^{\dag}(k) \\
                   \end{array}
                 \right)\left(
                          \begin{array}{cc}
                            D_{-}(k) & D_{+}(k) \\
                          \end{array}
                        \right),
\end{equation}
\begin{equation}\label{Dmpk}
    D_{\mp}(k)=N_j(k)T_{\mp}(k)+M_{j+}T_{\mp}(k-1)+M_{j-}T_{\mp}(k+1).
\end{equation}
Since $\mathcal N$ and $\mathcal R$ are Hermitian positively definite matrices, the characteristic equation $\det ({\mathcal R}-\lambda {\mathcal N})=0$ has positive roots specifying the generalized eigenvalues $\lambda_i, i=1,2,\ldots,\leq 8$. Let $\lambda_1$ and ${\mathcal Y}_1$ be the minimal eigenvalue and the corresponding generalized eigenvector, i.e., ${\mathcal R}{\mathcal Y}_1=\lambda_1 {\mathcal N}{\mathcal Y}_1$. This provides the values of $Y_{-1}$ and $Y_{1}$, which give the most accurate approximate solution at the prescribed values of $q_1$ and $q_4$. Substituting ${\mathcal Y}={\mathcal Y}_1$ in Eq.~(\ref{RZprime}) evaluates the fitness parameter $R=R(q_1,q_4)$ of this solution and thus makes possible to find the dispersion relations $q_4=q_4(q_1)$ and $q_1=q_1(q_4)$ as described below.

\subsubsection{Dispersion relations}
It follows from Eqs.~(\ref{field}) and (\ref{Om}) that the intensity $I_j$ of the plane harmonic wave $\textbf{A}_j e^{i \bm{K}_j \cdot \bm{x}}+\textbf{A}_j^\ast e^{-i \bm{K}_j \cdot \bm{x}}$ is specified by the dimensionless parameters $\Omega$ and $|\textbf{A}_j|^2$ as
\begin{equation}\label{intj}
    I_j=9.2962227\, \Omega^2|\textbf{A}_j|^2\times 10^{29} \frac{\text{W}}{\text{cm}^2}.
\end{equation}
The intensity parameter
\begin{equation}\label{IA}
    I_A = 2\sum_{j=1}^6 \left|\bm{A}_j \right|^2
\end{equation}
of the electromagnetic lattice $\textbf{A}^\prime$~(\ref{field}) plays an important role in the dispersion relations for various ESTCs~\cite{ESTCp3,ESTCp4}. In Ref.~\cite{Narozhny2004}, the probability of electron-positron pair creation by a focused laser pulse was calculated. It was shown that this process starts playing a role at intensities of the order of $10^{27} \text{W}/\text{cm}^2$. Such QED effects can be neglected at intensities treated in the presented article. As an example, let us consider the 2D-ESTC with parameters $L=-1$ and $g=1$ in Eq.~(\ref{A1A4}), hence $p=s, j=1, n=4$ in Eqs.~(\ref{dZdXj})--(\ref{Dmpk}), and $I_A=4(a_{12}^2+a_{42}^2)$. At $a_{12}=a_{42}$ and $I_A=0.0064$, the parameter $\Omega=0.1$ specifies the x-ray lattice with the wavelength $\lambda_0=2.426310\times 10^{-11}$ m, created by two circularly polarized waves with intensities $I_1=I_2=1.487395 \times 10^{25} \text{W}/\text{cm}^2$.

\begin{figure}
\includegraphics{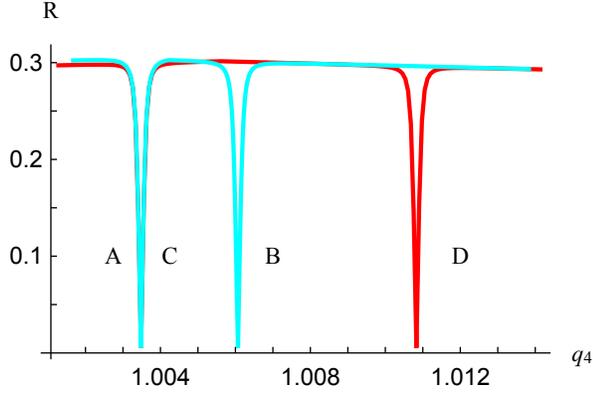}
\caption{\label{fig1q4ACBD}Fitness parameter $R$ against $q_4$: (A) line 1 for $s=-1$, (B) line 2 for $s=-1$, (C) line 1 for $s=1$, and (D) line 2 for $s=1$; $L=-1$; $g=1$; $\Omega=0.1$; $a_{12}=a_{42}$; $I_A=0.0064$; $q_1=0.024$; $k_m=3$.}
\end{figure}

The plots of function $R=R(s,q_4)$ in the form of two spectral curves of approximate solutions for $s=-1$ and $s=1$ at fixed $q_1=0.024$ are shown  in Fig.~\ref{fig1q4ACBD}. In this article, we consider only the positive frequency solutions ($q_4>0$). Each curve has two domains (``valeys'')  called spectral line 1 and line 2, where $R$ reaches local minima $q_{4}$ and $q'_{4}$, respectively. The width of these valleys is rapidly decreasing function of $k_m$ hence we use $k_m=3$ only for illustrative purposes in Fig.~\ref{fig1q4ACBD}, but in other cases we set $k_m \leq 14$ to obtain approximate particular solutions satisfying the fitness condition $R<10^{-17}$, whose deviations from the corresponding exact solutions are negligibly small. As $|k|$ increases, norms $|Y_k|$ of the Fourier amplitudes $Y_k$ in Eqs.~(\ref{YYk}) and (\ref{Zprime}) constitute a decreasing sequence which begins with $|Y_0|\approx 1$ and tends to zero. The rate of decrease becomes greater as $I_A$ is reduced, for an example, at $\Omega=0.1$ and $q_1=0$, for $I_A=0.0256$, $0.0064$, and $6.25\times 10^{-6}$, the condition $|Y_k|<10^{-16}$ is satisfied for $k>18$, $14$, and $6$, respectively. Some further examples for various types of ESTCs and values of $I_A$ and $\Omega$ are given in Refs.~\cite{ESTCp3,ESTCp4,ESTCp5}, in particular, see in Ref.~\cite{ESTCp3} Tables I and II comparing results of different truncations and Fig.~1 illustrating the limiting case at $I_A\rightarrow 0$.
%%%%%%%%%%%%%%%%%%%%%%%%%%%%%figure1%%%%%%%%%%%%%%%

\begin{figure}
\includegraphics{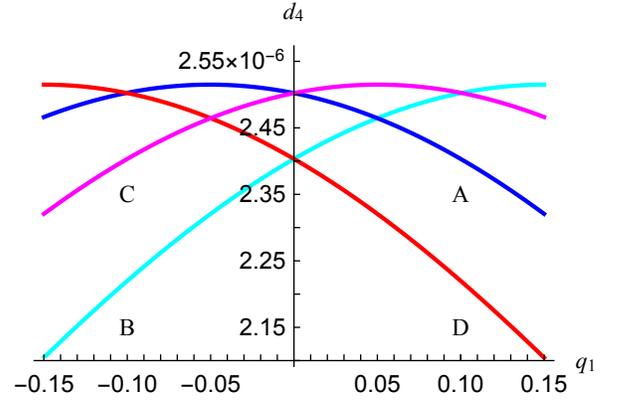}
\caption{\label{fig2q1d1d2}Function $d_4$ against $q_1$: (A) $d_4=d_{4-}(q_1)$, (B) $d_4=d'_{4-}(q_1)$, (C) $d_4=d_{4+}(q_1)$, and (D) $d_4=d'_{4+}(q_1)$;  $L=-1$; $g=1$; $\Omega=0.1$; $a_{12}=a_{42}$; $I_A=0.0064$.}
\end{figure}

It follows from the results of computer simulation that the dispersion relations can be written in terms of the representation $(1+d)^2 q_4^2=1+I_A+q_1^2$  as
\begin{eqnarray}\label{dispersion}
  q_{4\pm}&=&(1+d_{4\pm})^{-1}\sqrt{1+I_A+q_1^2}, \nonumber\\
  q'_{4\pm}&=&(1+d'_{4\pm})^{-1}\sqrt{1+I_A+(q_1 \pm\Omega)^2},
\end{eqnarray}
where $q_{4\pm}$ and $q'_{4\pm}$ are the solutions of the dispersion equations for $s=\pm 1$, specified by the positions of minima for lines 1 and 2, respectively. The functions $d_{4\pm}$ and $d'_{4\pm}$ are small and depend only weakly on $q_1$; see Fig.~\ref{fig2q1d1d2}. They vanish at $I_A=0$, when Eqs.~(\ref{dispersion}) reduce to the free-space dispersion relations for given $q_1$ and $q'_{1\pm}=q_1\pm \Omega$. For preliminary localization of spectral lines, one can substitute $d_{4\pm}=d'_{4\pm}=0$ in Eqs.~(\ref{dispersion}).

%%%%%%%%%%%%%%%%%%%%%%%%%%%%%figure2%%%%%%%%%%%%%%%
%%%%%%%%%%%%%%%%%%%%%%%%%%%%%%%%%%%%%%%%%%%%%%%%%%

\begin{figure}
\includegraphics{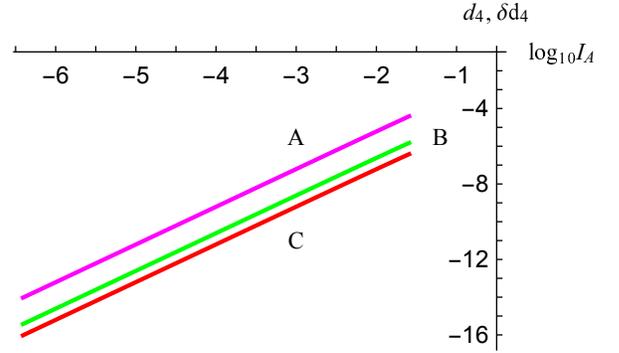}
\caption{\label{fig3abd1d2}Function $d_4=d_4(I_A,\Omega)$ and difference $\delta d_4=\delta d_4(I_A)$ against $\log_{10}I_A$: (A) $d_4=d_{4+}(I_A,0.1)$, (B) $\delta d_4=d_{4+}(I_A,0.1)-d'_{4+}(I_A,0.1)$, (C) $\delta d_4=d_{4+}(I_A,0.01)-d_{4+}(I_A,0.1)$;  $L=-1$; $g=1$; $a_{12}=a_{42}$; $q_1=0$.}
\end{figure}

When the particular solutions are found with the required accuracy, the parameters
\begin{eqnarray}\label{d4pm}
  d_{4\pm}&=&-1+(q_{4\pm})^{-1}\sqrt{1+I_A+q_1^2}, \nonumber\\
  d'_{4\pm}&=&-1+(q'_{4\pm})^{-1}\sqrt{1+I_A+(q_1 \pm\Omega)^2}
\end{eqnarray}
provide a convenient graphic description of the dispersion interrelations; see Figs.~\ref{fig2q1d1d2} and \ref{fig3abd1d2}. In particular,  it follows from Fig.~\ref{fig2q1d1d2} and Eqs.~(\ref{dispersion}) that $d_{4-}=d_{4+}\neq d'_{4-}=d'_{4+}$ and hence $q_{4-}=q_{4+}\neq q'_{4-}=q'_{4+}$ at $q_1=0$. Although lines A and C appear coinciding in Fig.~\ref{fig1q4ACBD}, their minima do not coincide since $d_{4-}\neq d_{4+}$ and $q_{4-}\neq q_{4+}$ at $q_1=0.024$. The differences $|d_{4+}-d_{4-}|$ and $|d'_{4+}-d'_{4-}|$ are small, Fig.~\ref{fig3abd1d2} illustrates the dependence of $d_{4+}$ and $d'_{4+}$ on $I_A$ for two values of the lattice frequency $\Omega$.

%%%%%%%%%%%%%%%%%%%%%%%%%%%%%figure3%%%%%%%%%%%%%%%
%%%%%%%%%%%%%%%%%%%%%%%%%%%%%%%%%%%%%%%%%%%%%%%%%%

\subsubsection{Integrals of motion and mean values of operators}
Let us impose the normalization condition
\begin{equation}\label{normcond}
    \langle\Psi^\dag\Psi\rangle\equiv\frac{1}{2\pi}\int_0^{2\pi}\Psi^{\dag}\Psi d\varphi_j = \sum_{k=-\infty}^{+\infty}Y_k^{\dag}Y_k=1
\end{equation}
on the electron wave functions $\Psi$ Eq.~(\ref{psiq1q4}) of both types $\{L,j,n\}=\{-1,4,1\}$ and $\{1, 1, 4\}$. It follows from Eqs.~(\ref{VWX4}) and (\ref{dVWdX4}) that $d(\Psi^{\dag}\Psi)/d\varphi_4=0$, and hence
\begin{equation}\label{YdagY}
    \Psi^{\dag}\Psi = \sum_{k=-\infty}^{+\infty}Y_k^{\dag}Y_k=1
\end{equation}
is the integral of motion at $L=-1$. Similarly, it follows from Eqs.~(\ref{VWX1}) and (\ref{dVWdX1}) that $d(\Psi^{\dag}\alpha_{1}\Psi)/d\varphi_1=0$, and hence
\begin{equation}\label{v1YdagY}
   \Psi^{\dag}\alpha_{1}\Psi = p\sum_{k=-\infty}^{+\infty}Y_k^{\dag}\left(
                                                                                  \begin{array}{cccc}
                                                                                    0 & -1 & 0 & 0 \\
                                                                                    -1 & 0 & 0 & 0 \\
                                                                                    0 & 0 & 0 & 1 \\
                                                                                    0 & 0 & 1 & 0 \\
                                                                                  \end{array}
                                                                                \right)
   Y_k
\end{equation}
is the integral of motion at $L=1$.

The Hermitian forms for the operators of probability current density (velocity) $j_k=c\alpha_k$ and spin $S_k=\frac{\hbar}{2}\Sigma_k$ with respect to $\Psi$ Eq.~(\ref{psiq1q4}) result in the vector fields $\textbf{j}=c\textbf{v}$ and $\textbf{S}=\frac{\hbar}{2}\textbf{s}$, where
\begin{eqnarray}\label{ve123}
  \textbf{v}&=&\sum_{k=1}^3 \textbf{e}_k (\Psi^\dag \alpha_k\Psi)\nonumber\\
  &=&v_1\textbf{e}_1+2v_0(\textbf{e}_2 p\cos\Phi-\textbf{e}_3\sin\Phi),
\end{eqnarray}
\begin{eqnarray}\label{v1v0}
  v_1&=&2p \rm{Re}(-y_1^{*}y_2+y_3^{*}y_4),\nonumber \\
  v_0&=&|y_1^{*}y_4+y_2^{*}y_3|,\nonumber \\
  \Phi&=&\delta + (-1)^{j}s\varphi_n,\quad e^{i\delta}=(y_1^{*}y_4+y_2^{*}y_3)/v_0,
\end{eqnarray}
\begin{eqnarray}\label{se123}
  \textbf{s}&=&\sum_{k=1}^3 \textbf{e}_k (\Psi^\dag \Sigma_k\Psi)\nonumber\\
  &=&s_1\textbf{e}_1+2s_0(\textbf{e}_2 p\cos\Phi'-\textbf{e}_3\sin\Phi'),
\end{eqnarray}
\begin{eqnarray}\label{s1s0}
  s_1&=&p(-|y_1|^2-|y_2|^2+|y_3|^2+|y_4|^2),\nonumber\\
  s_0&=&|y_1^{*}y_3+y_2^{*}y_4|,\nonumber \\
  \Phi'&=&\delta' + (-1)^{j}s\varphi_n,\quad e^{i\delta'}=(y_1^{*}y_3+y_2^{*}y_4)/s_0.
\end{eqnarray}
Here, $v_0, v_1, \delta,$ and  $s_0, s_1, \delta'$ are periodic functions in $\varphi_j$. However, at $L=1$, the velocity component $v_1=\Psi^{\dag}\alpha_{1}\Psi$ Eq.~(\ref{v1v0}) is independent of $\varphi_1$ owing to Eq.~(\ref{v1YdagY}).

The mean values of Hamiltonian
\begin{equation}\label{Hamilton}
    H=c\sum_{k=1}^3\alpha_k p_k + m_e c^2\alpha_4,
\end{equation}
operators of kinetic momentum
\begin{equation}\label{pk}
   p_k=-i\hbar\frac{\partial}{\partial x_k} - \frac{e}{c}A_k,
\end{equation}
velocity  $j_k$, and spin $S_k$ with respect to the wave function $\Psi$~(\ref{psiq1q4}) can be conveniently expressed in terms of $4\times 4$ matrix
\begin{equation}\label{matC}
    C=\langle Y Y^\dag \rangle=\sum_{-\infty}^{+\infty}Y_{l} Y_{l}^{\dag}
\end{equation}
with the unit trace and components $C_{lm}=\langle y_l y_m^{*}\rangle; l, m=1, 2, 3, 4$. Because of the foregoing normalization condition, the mean value $\langle L \rangle$ of a linear operator $L$ with respect to the wave function $\Psi$ reduces to the mean value of the corresponding Hermitian form:
 \begin{equation}\label{meanL0}
    \langle L \rangle=\frac{1}{4\pi^2}\int_0^{2\pi}d\varphi_1\int_0^{2\pi}d\varphi_4\Psi^{\dag} L \Psi.
 \end{equation}
The mean values $\langle j_k \rangle$, $\langle p_k \rangle$, and $\langle S_k \rangle$ are zero at $k=2, 3$ for both types of wave function, whereas the nonzero normalized mean values of the velocity $V_1$, the momentum $P_1$, and the spin $\Sigma_{10}$ are defined by the relations
\begin{equation}\label{mV1}
    V_1=\langle j_1 \rangle/c=2p \rm{Re}(C_{43}-C_{21}) \text{ for } L=\pm 1,
\end{equation}
\begin{eqnarray}\label{mP1}
  P_1&=&\langle p_1 \rangle/(m_e c)=q_1+s\Omega(C_{33}+C_{44}) \text{ for } L=-1,\nonumber\\
  &=&q_1+\Omega\sum_{k=-\infty}^{+\infty}k Y_k^{\dag}Y_k \text{ for } L=1,
\end{eqnarray}
\begin{equation}\label{mS1}
    \langle S_1 \rangle = \frac{\hbar}{2}\Sigma_{10},\quad \Sigma_{10} = p(-C_{11}-C_{22}+C_{33}+C_{44})
\end{equation}
for $L=\pm 1$.
The normalized energy $E$ is given by
\begin{eqnarray}\label{mE}
  E&=&\langle H \rangle/(m_e c^2)=\langle\frac{i}{2\pi} U\frac{\partial}{\partial X_4}\rangle\nonumber \\
   &=&q_4-\Omega\sum_{k=-\infty}^{+\infty} k Y_k^{\dag}Y_k \text{ for } L=-1,\nonumber \\
  &=&q_4+s\Omega(C_{33}+C_{44})\text{ for } L=1.
\end{eqnarray}

\begin{figure}
\includegraphics{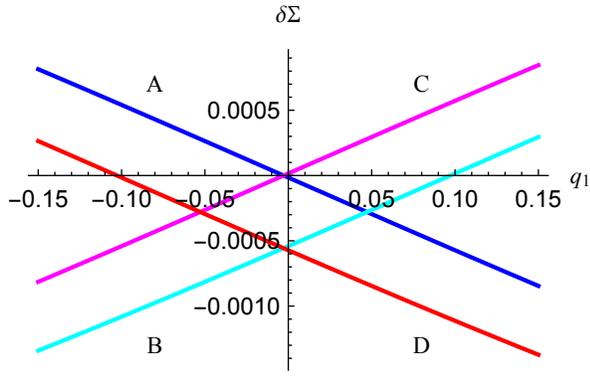}
\caption{\label{fig4dsigma1}Deviation $\delta \Sigma$ against $q_1$: (A) $\delta \Sigma =\Sigma_{10}-\Sigma_{1m}$ for $q_{4-}$ solution, (B) $\delta \Sigma =\Sigma_{10}+\Sigma_{1m}$ for $q'_{4-}$ solution, (C) $\delta \Sigma =\Sigma_{10}+\Sigma_{1m}$ for $q_{4+}$ solution, and (D) $\delta \Sigma =\Sigma_{10}-\Sigma_{1m}$ for $q'_{4+}$ solution;  $L=-1$; $g=1$; $\Omega=0.1$; $a_{42}/a_{12}=0.25$; $I_A=0.0064$.}
\end{figure}

For the particular solutions defined by Eqs.~(\ref{dispersion}) and illustrated in Figs. \ref{fig1q4ACBD} and \ref{fig2q1d1d2}, the mean values of spin are independent of $q_1$ and take the following values: $\langle S_1 \rangle=\frac{\hbar}{2}\Sigma_{1m}$ for $q_{4-}$ and $q'_{4+}$ solution, $\langle S_1 \rangle=-\frac{\hbar}{2}\Sigma_{1m}$ for $q'_{4-}$ and $q_{4+}$ solution, where $\Sigma_{1m}=|\Sigma_{10}|=0.99366079$. However, if $a_{12}\neq a_{42}$, then these mean values depend on $q_1$ and deviate from the above-listed values as shown in Fig.~\ref{fig4dsigma1}.

\begin{figure}
\includegraphics{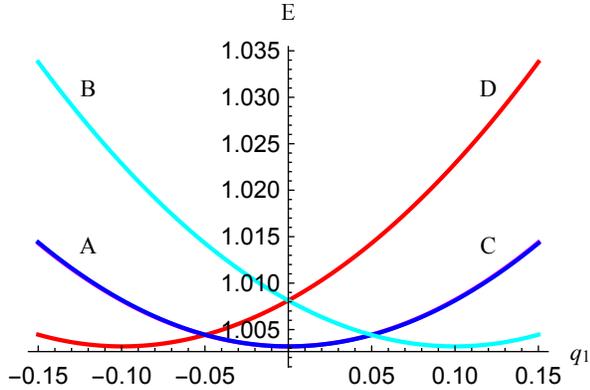}
\caption{\label{fig5q1E1E2}Energy $E$ against $q_1$: (A) $q_{4-}$ solution, (B) $q'_{4-}$ solution, (C) $q_{4+}$ solution, and (D) $q'_{4+}$ solution; $L=-1$; $g=1$; $\Omega=0.1$; $a_{42}=a_{12}$; $I_A=0.0064$. Curves A and C appear coinciding here, their difference is illustrated in Fig.~\ref{fig6q1E1mp}.}
\end{figure}

\begin{figure}
\includegraphics{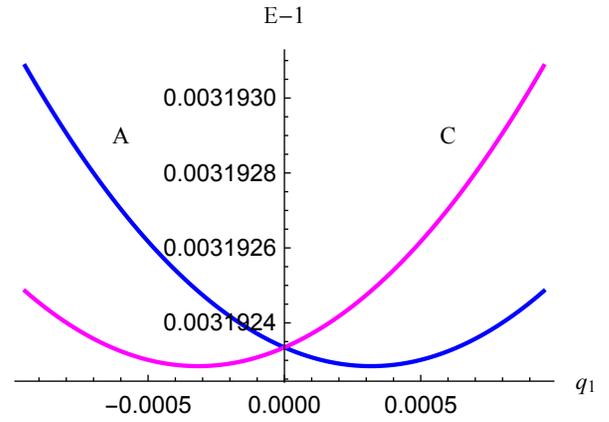}
\caption{\label{fig6q1E1mp}Energy $E$ against $q_1$: (A) $q_{4-}$ solution,  (C) $q_{4+}$ solution; $L=-1$; $g=1$; $\Omega=0.1$; $a_{42}=a_{12}$; $I_A=0.0064$.}
\end{figure}

\begin{figure}
\includegraphics{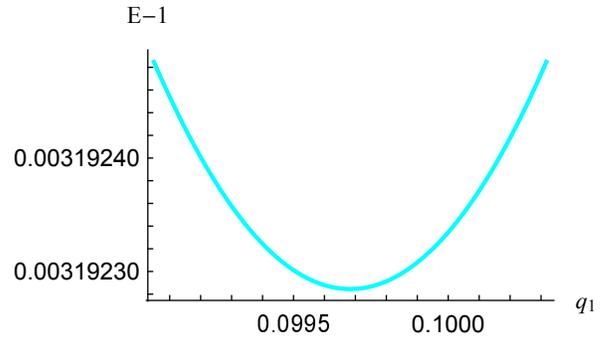}
\caption{\label{fig7q1E2m}Energy $E$ against $q_1$ for $q'_{4-}$ solution; $L=-1$; $g=1$; $\Omega=0.1$; $a_{42}=a_{12}$; $I_A=0.0064$.}
\end{figure}

Figures \ref{fig5q1E1E2}--\ref{fig7q1E2m} illustrate the dependence of energy $E$ on $q_1$ for these four solutions. The mean values of momentum $P_1$ linearly depend on $q_1$ and can be written as $P_1=q_1 \pm p_{10}$ and $P_1=q_1 \pm(\Omega-p_{10})$ for $q_{4\pm}$ and $q'_{4\pm}$ solutions, respectively, where $p_{10}=0.00031696$. They vanish at the minimum point of the corresponding energy function $E=E(q_1)$. The minimum energy value $E_{\rm min}=1.00319228$ is the same for all these four functions.

It follows from Eqs.~(\ref{dVWdX4}) and (\ref{abcd4}) that the same wave function $\Psi$ Eq.~(\ref{VWX4}) can be specified by two different sets of parameters $\{s,q_1\}$  and $\{\breve{s},\breve{q}_1\}=\{-s,q_1+s\Omega\}$. Similarly, from Eqs.~(\ref{dVWdX1}) and (\ref{abcd1}) follows that the same wave function $\Psi$ Eq.~(\ref{VWX1}) can be specified by both the sets of parameters $\{s,q_4\}$  and $\{\breve{s},\breve{q}_4\}=\{-s,q_4+s\Omega\}$. The corresponding amplitude functions are related as $\breve{V}=iW$ and $\breve{W}=-iV$. Figures \ref{fig5q1E1E2}--\ref{fig7q1E2m} illustrates in terms of energy $E$ the relation of (A) $q_{4-}$ solution with (D) $q'_{4+}$ solution and the relation of (C) $q_{4+}$ solution with (B) $q'_{4-}$ solution.

%%%%%%%%%%%%%%%%%%%%%%%%%%%%%figure4%%%%%%%%%%%%%%%
%%%%%%%%%%%%%%%%%%%%%%%%%%%%%%%%%%%%%%%%%%%%%%%%%%

%%%%%%%%%%%%%%%%%%%%%%%%%%%%%figure5%%%%%%%%%%%%%%%
%%%%%%%%%%%%%%%%%%%%%%%%%%%%%%%%%%%%%%%%%%%%%%%%%%

%%%%%%%%%%%%%%%%%%%%%%%%%%%%%figure6%%%%%%%%%%%%%%%
%%%%%%%%%%%%%%%%%%%%%%%%%%%%%%%%%%%%%%%%%%%%%%%%%%

%%%%%%%%%%%%%%%%%%%%%%%%%%%%%figure7%%%%%%%%%%%%%%%
%%%%%%%%%%%%%%%%%%%%%%%%%%%%%%%%%%%%%%%%%%%%%%%%%%

\section{\label{sec:local}Localized solutions}
\subsection{\label{sec:local4D}4D-ESTC}
Let $(u_k)$ be a set of complex scalar functions on a real manifold $\cal B$, satisfying the orthonormality condition
\begin{equation}\label{ortho}
    \int_{\cal B}u^{*}_k u_ld{\cal B}=\delta_{kl},
\end{equation}
where $d{\cal B}$ is the infinitesimal element of $d{\cal B}$, $u^{*}_k$ is the complex conjugate function to $u_k$, and $\delta_{kl}$ is the Kronecker symbol. Let us consider a superposition of particular solutions $\Psi$ Eq.~(\ref{sol1}) (termed below the ``beam" for the sake of brevity),
\begin{equation}\label{beam}
    \Psi_k=\int_{\xi_1}^{\xi_2}d\xi \int_{\cal B}d{\cal B} \nu u_k \sum_{n \in \mathcal L}c_n e^{i \bm{x}\cdot\bm{K}_n},
\end{equation}
where $\bm{K}_n=\bm{K}+\bm{G}_n=\kappa_e(\textbf{q},i q_4)+k_0 (\textbf{n},i n_4)$, the function $\textbf{q}=\textbf{q}(\xi,b)$ defines $q_4=q_4(\xi,b)$ owing to the dispersion relation $q_4=q_4(\textbf{q})$, and $c_n=c(n,\xi,b)$ are the bispinor Fourier amplitudes for given real $\xi\in [\xi_1,\xi_2]$ and $b\in{\cal B}$. The set of particular solutions forming the beam (beam base) is specified by functions $\bm{K}_n$ and $c_n$. The beam states are prescribed by the set of complex functions $u_k=u_k(b)$ and a real function $\nu=\nu(\xi,b)$ which is used below to obtain orthonormal beam sets. By setting the beam manifold, the beam base, and the beam states in various ways, one can obtain a multitude of localized solutions to the Dirac equation.

Let ${\cal B}$ be a two-dimensional manifold with the local coordinates $b=\{b_1, b_2\}$, and $d{\cal B}=g(b)db_1 db_2$. To find the function $\nu$, we use the following designations:
\begin{eqnarray}
  {\cal I}_3[f]&=&\int_{R^3}f(X_1,X_2,X_3,X_4) d X_1 d X_2 d X_3,\nonumber\\
  {\cal I}_4[f]&=&\Omega\int_0^{1/\Omega}f(X_1,X_2,X_3,X_4) d X_4,\nonumber\\
  {\cal I}_{34}[f]&=&{\cal I}_3[{\cal I}_4[f]].
\end{eqnarray}
Let us assume that a Hermitian operator ${\cal O}$ has the restriction ${\cal O}_n$ to the function $\exp(i \bm{x}\cdot\bm{K}_n)$, independent of $\bm{x}$ and defined by the relation
\begin{equation}\label{On}
    {\cal O}e^{i \bm{x}\cdot\bm{K}_n}=e^{i \bm{x}\cdot\bm{K}_n}{\cal O}_n.
\end{equation}
It follows from Eqs. (\ref{beam})--(\ref{On}) that
\begin{eqnarray}\label{I34a}
  {\cal I}_{34}\left[\Psi_k^{\dag}{\cal O}\Psi_l\right]&=&\int_{\xi_1}^{\xi_2}d\xi \int_{\cal B}d{\cal B} \nu u^{*}_k \sum_{m \in \mathcal L}c^{\dag}_m \nonumber\\
   &\times& \int_{\xi_1}^{\xi_2}d\xi' \int_{\cal B}d{\cal B}' \nu' u'_l \sum_{n \in \mathcal L}{\cal O}'_n c'_n  \nonumber\\
   &\times& \delta[\textbf{q}'-\textbf{q}+\Omega(\textbf{n}-\textbf{m})] {\cal I}_4\left[e^{i\Phi_4}\right],
\end{eqnarray}
where
\begin{equation}\label{Phi4}
    \Phi_4=-2\pi X_4[q'_4-q_4+\Omega(n_4-m_4)].
\end{equation}
For simplicity sake let us preset the function $\textbf{q}=\textbf{q}(\xi,b)$ such that the argument of the Dirac delta function in Eq.~(\ref{I34a}) vanishes if and only if $\textbf{q}'=\textbf{q}$ and $\textbf{n}=\textbf{m}$. The condition $\textbf{q}'=\textbf{q}$ results in  $q'_4-q_4\equiv q_4(\textbf{q}')-q_4(\textbf{q})=0$ and hence ${\cal I}_4\left[\exp(i\Phi_4)\right]=\delta_{m_4 n_4}$. Because of this, upon integrating with respect to $\xi'$ and $b'$ we obtain
\begin{equation}\label{I34b}
    {\cal I}_{34}\left[\Psi_k^{\dag}{\cal O}\Psi_l\right]=\int_{\xi_1}^{\xi_2}d\xi \int_{\cal B}d{\cal B} u^{*}_k u_l \nu^2\frac{g}{J}\sum_{m \in \mathcal L}c^{\dag}_m {\cal O}_m c_m,
\end{equation}
where $J=\partial(q_1,q_2,q_3)/\partial(\xi,b_1,b_2)$ is the Jacobian determinant of the mapping $(\xi,b)\mapsto \textbf{q}$.

From Eqs. (\ref{ortho}) and (\ref{I34b}) follows that the function
\begin{equation}\label{nuO}
    \nu=\sqrt{\frac{J}{g \Delta\xi \sum_{m \in \mathcal L}c^{\dag}_m {\cal O}_m c_m}},
\end{equation}
where $\Delta\xi=\xi_2-\xi_1$, defines the orthonormal beam set satisfying the condition
\begin{equation}\label{nuOa}
    {\cal I}_{34}\left[\Psi_k^{\dag}{\cal O}\Psi_l\right]=\delta_{k l}.
\end{equation}
In particular, the function
\begin{equation}\label{nuU}
    \nu=\sqrt{\frac{J}{g \Delta\xi \sum_{m \in \mathcal L}c^{\dag}_m c_m}}
\end{equation}
gives the beam set satisfying the condition
\begin{equation}\label{nuUa}
    {\cal I}_{34}\left[\Psi_k^{\dag}\Psi_l\right]=\delta_{k l}.
\end{equation}
In this case, the mean value of operator $\cal O$ with respect to the beam $\Psi_k$ can be written as
\begin{eqnarray}\label{meanO}
  \langle{\cal O}\rangle&=&{\cal I}_{34}\left[\Psi_k^{\dag}{\cal O}\Psi_k\right] \nonumber\\
   &=&\frac{1}{\Delta\xi}\int_{\xi_1}^{\xi_2}d\xi \int_{\cal B}d{\cal B}|u_k|^2 \frac{\sum_{m \in \mathcal L}c^{\dag}_m {\cal O}_m c_m}{\sum_{m \in \mathcal L}c^{\dag}_m c_m}.
\end{eqnarray}

\subsection{\label{sec:free}Free space}
By way of illustration let us consider localized solutions of the Dirac equation in free space, defined by the spherical harmonics $Y_l^m$.

\subsubsection{Beam base}
For a plane wave function
\begin{equation}\label{planew}
    \Psi=\Psi_0 e^{2\pi i(\textbf{q}\cdot\textbf{R}-q_4 X_4)},\quad \Psi_0=\left(
                                                                             \begin{array}{c}
                                                                               v_0 \\
                                                                               w_0 \\
                                                                             \end{array}
                                                                           \right),
\end{equation}
Eq.~(\ref{diraceq}) reduces to the matrix equation
\begin{equation}\label{matD}
   D\Psi_0=0,\quad D=\left(
        \begin{array}{cc}
          (1-q_4)I & \textbf{q}\cdot\bm{\sigma} \\
          \textbf{q}\cdot\bm{\sigma} & -(1+q_4)I \\
        \end{array}
      \right),
\end{equation}
where
\begin{eqnarray}
                            I&=&\left(
                                 \begin{array}{cc}
                                   1 & 0 \\
                                   0 & 1 \\
                                 \end{array}
                               \right),\quad \textbf{q}\cdot\bm{\sigma}=\sum_{k=1}^{3}q_k\sigma_k, \nonumber\\
            \sigma_1&=& \left(
                 \begin{array}{cc}
                   0 & 1 \\
                   1 & 0 \\
                 \end{array}
               \right), \sigma_2= \left(
                 \begin{array}{cc}
                   0 & -i \\
                   i & 0 \\
                 \end{array}
               \right),  \sigma_3= \left(
                 \begin{array}{cc}
                   1 & 0 \\
                   0 & -1 \\
                 \end{array}
               \right).\nonumber
\end{eqnarray}
The nonzero solutions are described by the relation
\begin{equation}\label{w0v0}
    w_0=\frac{\textbf{q}\cdot\bm{\sigma}}{1+q_4} v_0,
\end{equation}
where $q_4=\sqrt{1+\textbf{q}^2}$ and the spinor $v_0$ may be prescribed arbitrarily.

The spherical harmonics $Y_l^m=Y_l^m(\vartheta,\varphi)$ satisfy the relations
\begin{equation}\label{normYlm}
    \int_0^{2\pi} d\varphi\int_0^\pi \sin{\vartheta}d\vartheta {Y_l^m}^*Y_{l'}^{m'}=\delta_{ll'}\delta_{mm'},
\end{equation}
i.e., the manifold ${\cal B}=S^2$ is a unit sphere, $d{\cal B}=\sin\vartheta d\vartheta d\varphi$, and $g=\sin\vartheta$ [see Eq.~(\ref{ortho})].

Since $q_4$ depend only on the magnitude $q=|\textbf{q}|$ of the vector $\textbf{q}=q\textbf{e}_q$, we specify its direction by spherical coordinates $\theta$ and $\phi$ as
\begin{equation}\label{qeq}
    \textbf{e}_q=\textbf{e}_1\cos\theta+\sin\theta(\textbf{e}_2\cos\phi+\textbf{e}_3\sin\phi).
\end{equation}
Taking into consideration Eqs.~(\ref{planew})--(\ref{w0v0}) and (\ref{qeq}), we define two linearly independent solutions for each given $\textbf{q}$ by amplitudes
\begin{equation}\label{Psip}
    \Psi_p=\frac{1}{N_0}\left(
                              \begin{array}{c}
                                v_p \\
                                p\varrho_{0} v_p \\
                              \end{array}
                            \right),
\end{equation}
where $p=-1, 1$, and
\begin{equation}\label{rhoN0}
    \varrho_0=\frac{q}{q_4+1},\quad N_0=\sqrt{1+\varrho_0^2},
\end{equation}
\begin{eqnarray}\label{vpvp0}
  v_p&=&\exp\left(-\frac{i\phi}{2}\sigma_1\right)\exp\left(-\frac{i\theta}{2}\sigma_3\right)\exp\left(p_h\frac{i\phi}{2}\sigma_1\right)v_p(0)\nonumber \\
  &=&v_p(0)\cos\frac{\theta}{2}\exp\left[p\frac{i\phi}{2}(p_h -1)\right]\nonumber\\
  &-&i v_{-p}(0)\sin\frac{\theta}{2}\exp\left[p\frac{i\phi}{2}(p_h +1)\right],\nonumber\\
  v_p(0)&=&\frac{1}{\sqrt{2}}\left(
                                \begin{array}{c}
                                  1 \\
                                  p \\
                                \end{array}
                              \right),
\end{eqnarray}
and a real coefficient $p_h$ may be set arbitrarily. The bispinor amplitudes $\Psi_p$ and the spinors $v_p$ satisfy the relations
\begin{eqnarray}\label{Psipvp}
  \Psi_p^\dag \Psi_p=v_p^\dag v_p&=&1,\quad  \Psi_p^\dag \Psi_{-p}=v_{p}^\dag v_{-p}=0,\nonumber\\
    (\textbf{e}_q\cdot\bm{\sigma})v_p&=&p v_p,
\end{eqnarray}
for any values of $\theta, \phi$, and $p_h$. The spinors $v_p$ and $v_{-p}$ are interrelated as $v_p(\theta+\pi,\phi+2\pi)=iv_{-p}(\theta,\phi)$ at $p_h=0$.

The function $\textbf{q}=\textbf{q}(\xi,b)$ can be prescribed by various mappings $\{\xi,\vartheta,\varphi\}\mapsto\{q,\theta,\phi\}$. In this article, let us consider the beams with $\xi=q, \theta=\chi\vartheta, \phi=\varphi$, where $\chi$ is some real parameter, $0<\chi\leq 1$. From Eqs.~(\ref{beam}), (\ref{nuU}), (\ref{planew}), and (\ref{qeq})--(\ref{Psipvp}) we obtain the localized solutions
\begin{equation}\label{Psilmp}
    \Psi_{l,p}^m=\int_{q_a}^{q_b}dq\int_0^{\pi}d\vartheta\nu\sin{\vartheta}\int_0^{2\pi}d\varphi Y_l^m \Psi_p e^{i\Phi},
\end{equation}
where
\begin{eqnarray}\label{nuqchi}
  \nu&=&\frac{q}{\sqrt{2\delta q}}\nu_{\chi},\quad  \nu_{\chi}=\sqrt{\frac{\chi\sin{\chi\vartheta}}{\sin\vartheta}},\nonumber\\
  \Phi&=&2\pi(q R_q -q_4X_4),\quad R_q=\textbf{R}\cdot\textbf{e}_q, \nonumber \\
  \delta q&=&(q_b -q_a)/2,\quad q_0=(q_a+q_b)/2.
\end{eqnarray}
They satisfy the orthonormality condition
\begin{equation}\label{ortholmp}
    {\cal I}_3\left[{\Psi_{l,p}^m}^{\dag}\Psi_{l',p'}^{m'}\right]=\delta_{ll'}\delta_{mm'}\delta_{pp'}.
\end{equation}

\subsubsection{Probability density and vortex currents}
The spherical harmonics are defined by the relations
\begin{eqnarray}\label{Ylm}
  Y_l^m&=&N_{lm}P_l^{|m|}(\cos\vartheta)e^{im\varphi}, \nonumber\\
  N_{lm}&=&\sqrt{\frac{(2l+1)(l-|m|)!}{4\pi(l+|m|)!}},
\end{eqnarray}
where $P_l^{|m|}$ is the associated Legendre function~\cite{Korn}. The phase factor $e^{im\varphi}$ in $Y_l^m$ and the spinors $v_p$ Eq.~(\ref{vpvp0}) preset the initial phases of plane waves in $\Psi_{l,p}^m$ Eq.~(\ref{Psilmp}) as functions of $\vartheta$ and $\varphi$. In particular, $v_p$ become periodic in $\varphi$ with the period $2\pi$ at $p_h=\pm 1$.

\begin{figure}
\includegraphics{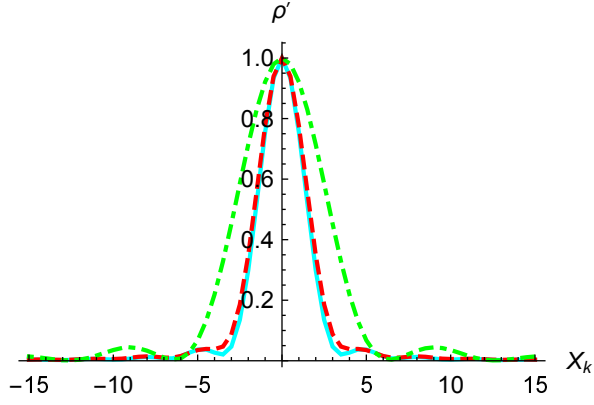}
\caption{\label{fig8pla001}Relative probability density $\rho'$ for $l=0, m=0$, and $p=\pm 1$ against $X_k$ along the coordinate axes $X_1$ (solid line), $X_2$ (dashed line), and $X_3$ (dash-and-dot line); $p_h=0$; $\chi=1$; $q_0=1$; ${\delta q}=10^{-8}$; $\rho_{00}=9.05507\times 10^{-8}$.}
\end{figure}

\begin{figure}
\includegraphics{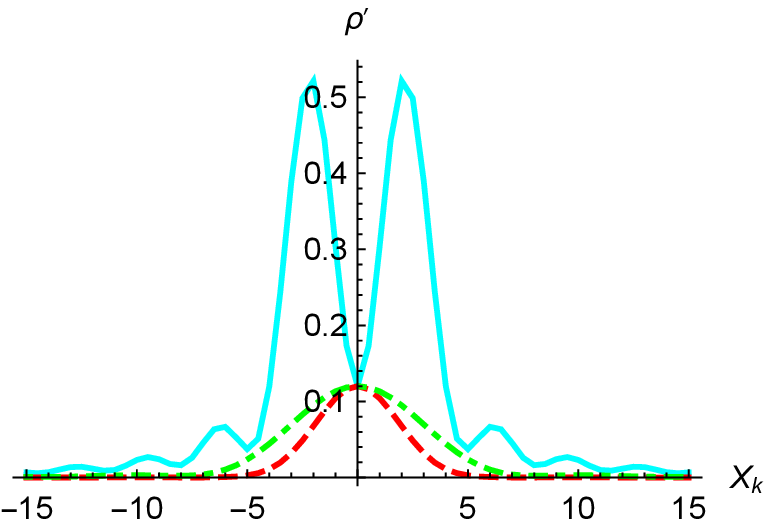}
\caption{\label{fig9pla101}Relative probability density $\rho'$ for $l=1, m=0$, and $p=\pm 1$ against $X_k, k=1, 2, 3$; the other parameters and the notations are the same as described in the caption of Fig.~\ref{fig8pla001}.}
\end{figure}

\begin{figure}
\includegraphics{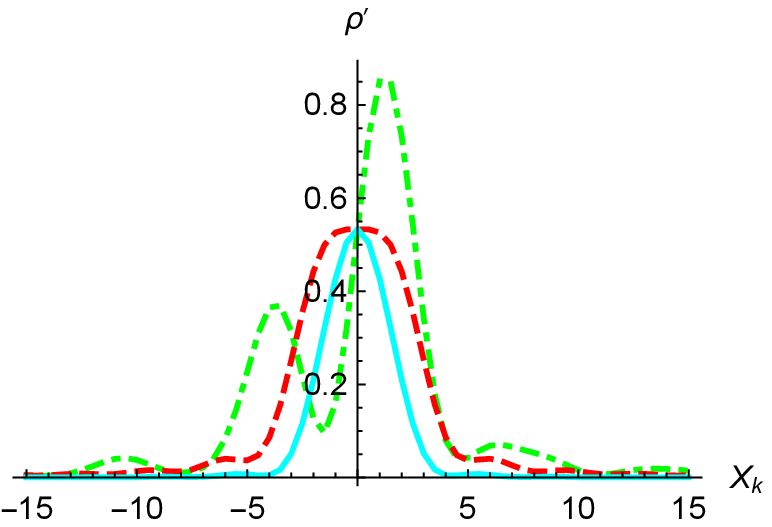}
\caption{\label{fig10pla111}Relative probability density $\rho'$ for $l=1, m=1$, and $p=\pm 1$  against $X_k, k=1, 2, 3$; the other parameters and the notations are the same as described in the caption of Fig.~\ref{fig8pla001}.}
\end{figure}

\begin{figure}
\includegraphics{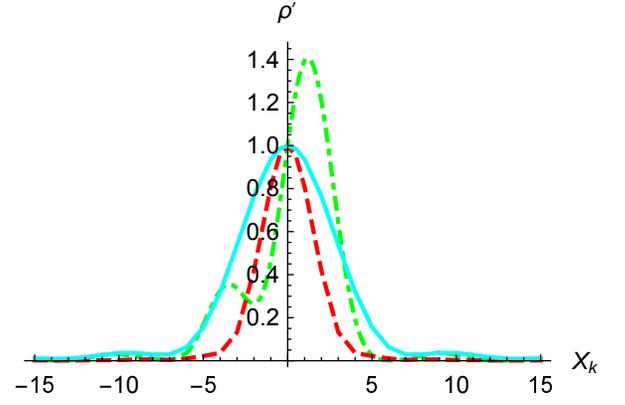}
\caption{\label{fig11pld001}Relative probability density $\rho'$ for $l=0, m=0$, and $p=-1$  against $X_k, k=1, 2, 3$;  $\chi=0.5$; $\rho_{00}=4.3949\times 10^{-8}$; the other parameters and the notations are the same as described in the caption of Fig.~\ref{fig8pla001}.}
\end{figure}

\begin{figure}
\includegraphics{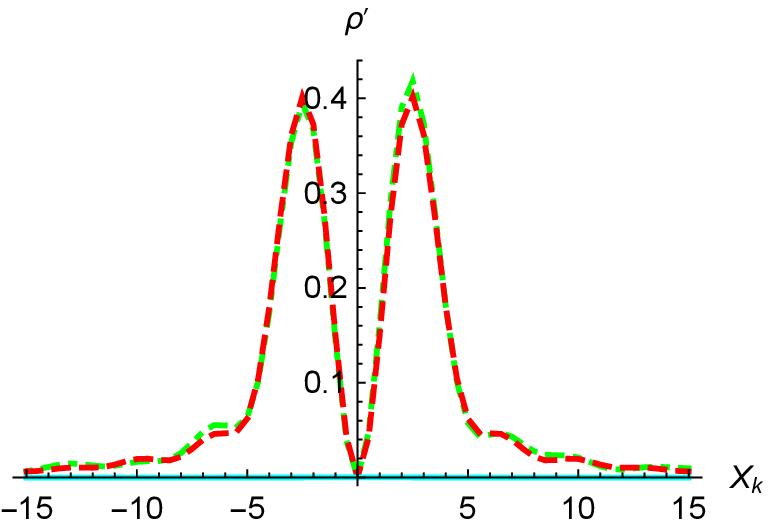}
\caption{\label{fig12plb111}Relative probability density $\rho'$ for $l=1, m=1$, and $p=-1$ against $X_k$; $p_h=-1$; $\rho_{00} = 1.0928\times 10^{-7}$; the other parameters and the notations are the same as described in the caption of Fig.~\ref{fig8pla001}.}
\end{figure}

\begin{figure}
\includegraphics{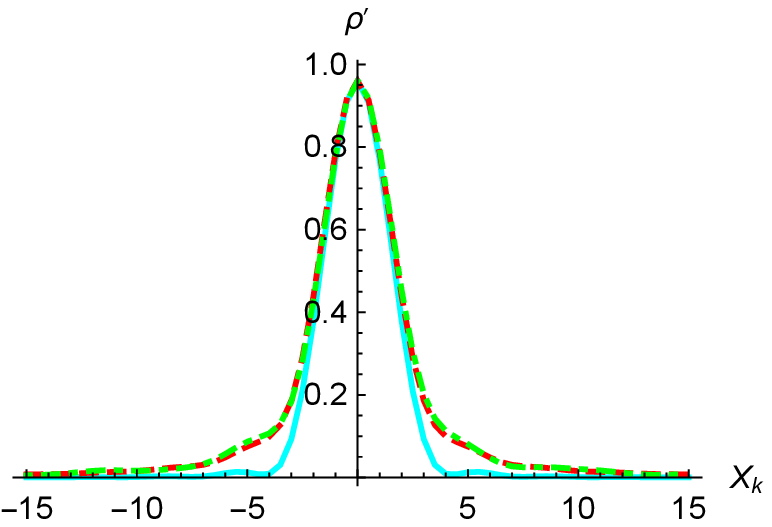}
\caption{\label{fig13plb112}Relative probability density $\rho'$ for $l=1, m=1$, and $p=1$ against $X_k$; $p_h=-1$; $\rho_{00} = 1.0928\times 10^{-7}$; the other parameters and the notations are the same as described in the caption of Fig.~\ref{fig8pla001}.}
\end{figure}

It is convenient to compare probability densities of beams defined by different spherical harmonics in terms of the relative density $\rho'=|\Psi_{l,p}^m|^2/\rho_{00}$, where $\rho_{00}=|\Psi_{0,p}^0|^2$ is the probability density of the beam $\Psi_{0,p}^0$ at the origin of coordinates $\textbf{x}=0$ (see Fig.~\ref{fig8pla001}). Figures \ref{fig8pla001}--\ref{fig13plb112} illustrate probability density for quasimonochromatic (${\delta q}\ll q_0$) three-dimensionally localized beams defined by some spherical harmonics at $p_h=0$ and $p_h=-1$ for $\chi =1/2$ and  $\chi =1$.

In the case $p_h=0$ and $\chi=1$, illustrated in Figs.~\ref{fig8pla001}--\ref{fig10pla111}, $\rho_{00}$ and $\rho'$ are the same for the both beam states $p=-1$ and $p=1$. The functions $\rho'=\rho'(l,m,p,X_k)$ are symmetric about the axis $X_k$ for $m=0$. This symmetry breaks at $m\neq 0$, in particular, for the $X_3$ axis at $l=m=1$ (see Fig.~\ref{fig10pla111}). The probability densities of beams $\Psi_{1,p}^{-1}$ and $\Psi_{1,p}^{1}$ are related as $\rho'(1,-1,p,X_3)=\rho'(1,1,p,-X_3)$.

The parameter $\chi$ relating the polar angle $\theta=\chi\vartheta$ in Eq.~(\ref{qeq}) with the angle $\vartheta$ in Eq.~(\ref{Ylm}) specifies, in particular, the solid angle $\Omega_q$ which embraces all propagation directions of the plane waves creating the beams. Figures \ref{fig8pla001} and \ref{fig11pld001} illustrate the probability density changes for the beam $\Psi_{0,1}^0$ in passing from $\chi=1, \Omega_q=4\pi$ to $\chi=0.5, \Omega_q=2\pi$. In the latter case, the beam states $p=-1$ and $p=1$ have different densities which are related as $\rho'(0,0,-1,X_3)=\rho'(0,0,1,-X_3)$ along the $X_3$ axis.

Figures \ref{fig10pla111}, \ref{fig12plb111}, and \ref{fig13plb112} illustrate the probability density changes for the beams $\Psi_{1,-1}^1$ and  $\Psi_{1,1}^1$ when in use $p_h=-1$ instead of $p_h=0$, i.e., the spinors $v_p$ become periodic in $\varphi$.

\begin{figure}
\includegraphics{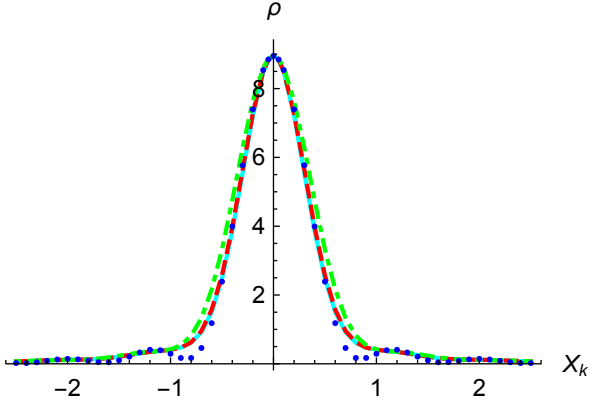}
\caption{\label{fig14plc001}Probability density $\rho$ for $l=0, m=0$, and $p=\pm 1$ against $X_k$ along the coordinate axes $X_1$ (solid line), $X_2$ (dashed line almost coincident with solid one), $X_3$ (dash-and-dot line), $X_4$ (dotted line); $p_h=0$; $\chi=1$; $q_0={\delta q}=1$.}
\end{figure}

One can obtain the beams localized with respect to all space-time coordinates by integrating over a wide range $\delta q$. The probability density for the four-dimensionally localized beams $\Psi_{0,p}^0$ is shown in Fig. \ref{fig14plc001}.

\begin{figure}
\includegraphics{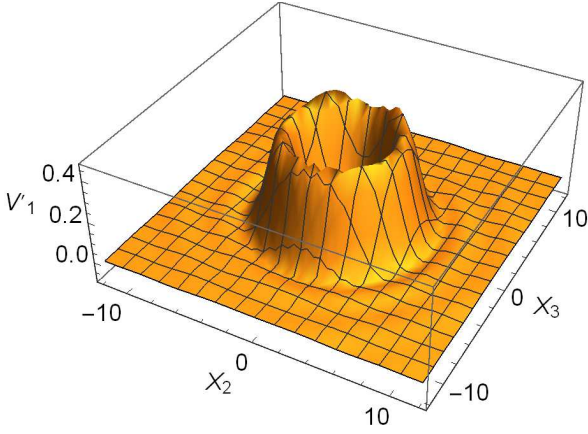}
\caption{\label{fig15v1x23ple}Component $V'_1$ of the probability current as a function of $X_2$ and $X_3$; $X_1=X_4=0$; $l=1, m=1, p=-1$; $p_h=-1$; $\chi=0.5$; $q_0=1$; ${\delta q}=10^{-8}$; $\rho_{00}=1.90536\times 10^{-8}$.}
\end{figure}

\begin{figure}
\includegraphics{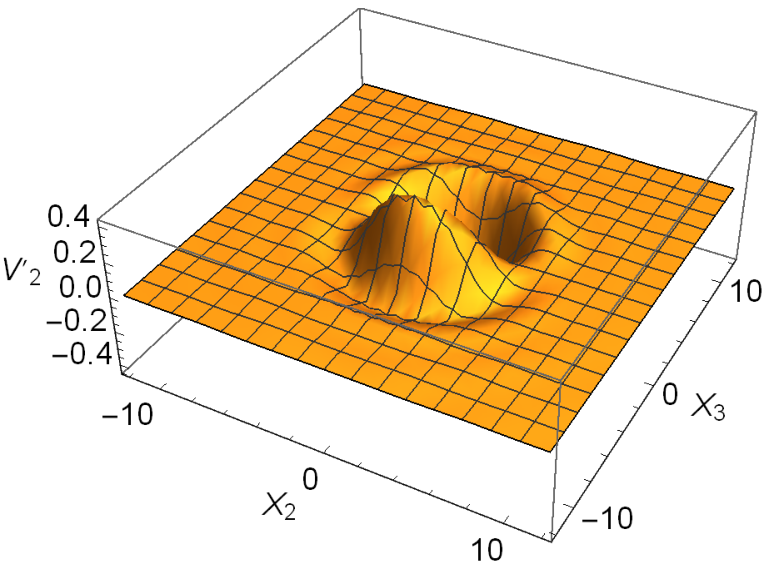}
\caption{\label{fig16v2x23ple}Component $V'_2$ of the probability current as a function of $X_2$ and $X_3$; the other parameters are the same as described in the caption of Fig.~\ref{fig15v1x23ple}.}
\end{figure}

\begin{figure}
\includegraphics{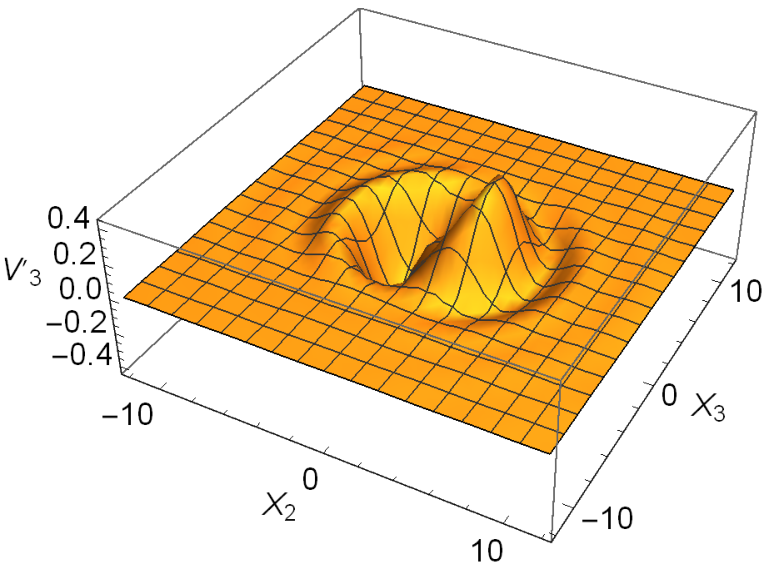}
\caption{\label{fig17v3x23ple}Component $V'_3$ of the probability current as a function of $X_2$ and $X_3$; the other parameters are the same as described in the caption of Fig.~\ref{fig15v1x23ple}.}
\end{figure}

\begin{figure}
\includegraphics{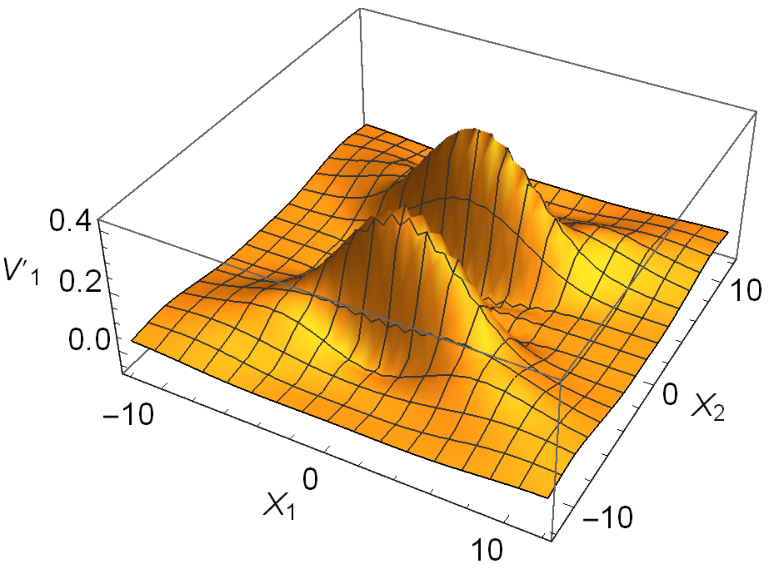}
\caption{\label{fig18v1x12ple}Component $V'_1$ of the probability current as a function of $X_1$ and $X_2$; $X_3=X_4=0$; the other parameters are the same as described in the caption of Fig.~\ref{fig15v1x23ple}.}
\end{figure}

\begin{figure}
\includegraphics{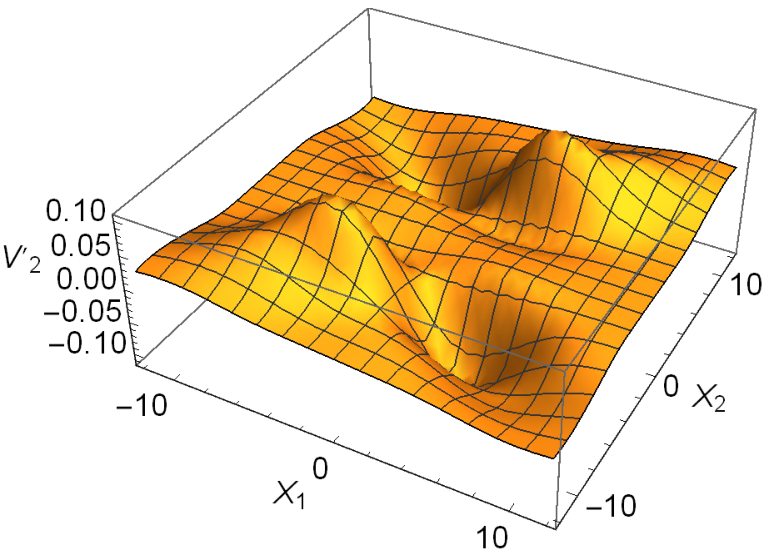}
\caption{\label{fig19v2x12ple}Component $V'_2$ of the probability current as a function of $X_1$ and $X_2$; $X_3=X_4=0$; the other parameters are the same as described in the caption of Fig.~\ref{fig15v1x23ple}.}
\end{figure}

\begin{figure}
\includegraphics{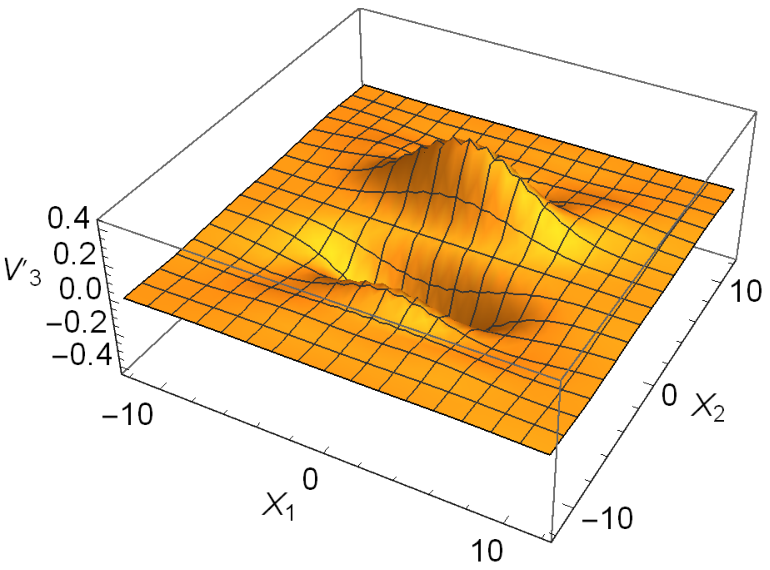}
\caption{\label{fig20v3x12ple}Component $V'_3$ of the probability current as a function of $X_1$ and $X_2$; $X_3=X_4=0$; the other parameters are the same as described in the caption of Fig.~\ref{fig15v1x23ple}.}
\end{figure}

The localized states defined by the spherical harmonics $\Psi=\Psi_{l,p}^m$ have a complex vortex structure of probability currents $v_k=\Psi^\dag \alpha_k\Psi, k=1, 2, 3$. Figures \ref{fig15v1x23ple}--\ref{fig20v3x12ple} illustrate these vortex currents for the beam $\Psi_{1,-1}^1$. To simplify these graphic representations, we use the normalized components $V'_k=v_k/\rho_{00}$.

%%%%%%%%%%%%%%%%%%%%%%%%%%%%%figure8%%%%%%%%%%%%%%%
%%%%%%%%%%%%%%%%%%%%%%%%%%%%%%%%%%%%%%%%%%%%%%%%%%

%%%%%%%%%%%%%%%%%%%%%%%%%%%%%figure9%%%%%%%%%%%%%%%
%%%%%%%%%%%%%%%%%%%%%%%%%%%%%%%%%%%%%%%%%%%%%%%%%%

%%%%%%%%%%%%%%%%%%%%%%%%%%%%%figure10%%%%%%%%%%%%%%%
%%%%%%%%%%%%%%%%%%%%%%%%%%%%%%%%%%%%%%%%%%%%%%%%%%

%%%%%%%%%%%%%%%%%%%%%%%%%%%%%figure11%%%%%%%%%%%%%%%
%%%%%%%%%%%%%%%%%%%%%%%%%%%%%%%%%%%%%%%%%%%%%%%%%%

%%%%%%%%%%%%%%%%%%%%%%%%%%%%%figure12%%%%%%%%%%%%%%%
%%%%%%%%%%%%%%%%%%%%%%%%%%%%%%%%%%%%%%%%%%%%%%%%%%

%%%%%%%%%%%%%%%%%%%%%%%%%%%%%figure13%%%%%%%%%%%%%%%
%%%%%%%%%%%%%%%%%%%%%%%%%%%%%%%%%%%%%%%%%%%%%%%%%%

%%%%%%%%%%%%%%%%%%%%%%%%%%%%%figure14%%%%%%%%%%%%%%%
%%%%%%%%%%%%%%%%%%%%%%%%%%%%%%%%%%%%%%%%%%%%%%%%%%

%%%%%%%%%%%%%%%%%%%%%%%%%%%%%figure15%%%%%%%%%%%%%%%
%%%%%%%%%%%%%%%%%%%%%%%%%%%%%%%%%%%%%%%%%%%%%%%%%%

%%%%%%%%%%%%%%%%%%%%%%%%%%%%%figure16%%%%%%%%%%%%%%%
%%%%%%%%%%%%%%%%%%%%%%%%%%%%%%%%%%%%%%%%%%%%%%%%%%

%%%%%%%%%%%%%%%%%%%%%%%%%%%%%figure17%%%%%%%%%%%%%%%
%%%%%%%%%%%%%%%%%%%%%%%%%%%%%%%%%%%%%%%%%%%%%%%%%%

%%%%%%%%%%%%%%%%%%%%%%%%%%%%figure18%%%%%%%%%%%%%%%
%%%%%%%%%%%%%%%%%%%%%%%%%%%%%%%%%%%%%%%%%%%%%%%%%

%%%%%%%%%%%%%%%%%%%%%%%%%%%%%figure19%%%%%%%%%%%%%%%
%%%%%%%%%%%%%%%%%%%%%%%%%%%%%%%%%%%%%%%%%%%%%%%%%%

%%%%%%%%%%%%%%%%%%%%%%%%%%%%%%figure20%%%%%%%%%%%%%%%
%%%%%%%%%%%%%%%%%%%%%%%%%%%%%%%%%%%%%%%%%%%%%%%%%%

\subsubsection{\label{sec:mean}Mean values}
The mean value of the spin operator \textbf{S} with respect to the wave function $\Psi_{l,p}^m$ Eq.~(\ref{Psilmp}) is given by
\begin{equation}\label{mspin}
    \langle S_1\rangle=\frac{\hbar}{2}\langle \Sigma_1\rangle= \frac{p\hbar}{2}Z_l^{|m|}(\chi), \langle S_2\rangle=\langle S_3\rangle=0,
\end{equation}
where
\begin{equation}\label{Zlm}
    Z_l^{|m|}(\chi)=2\pi\int_0^\pi |Y_l^m|^2 \cos{\chi\vartheta} \sin\vartheta d\vartheta.
\end{equation}
The operator of the orbital angular momentum
\begin{equation}\label{orbital}
    \textbf{L}=-i\hbar \textbf{r}\times\frac{\partial}{\partial\textbf{r}}
\end{equation}
has the mean value given by
\begin{equation}\label{morbital}
    \langle L_1\rangle=\hbar\left[m'-\frac{p}{2}Z_l^{|m|}(\chi)\right], \langle L_2\rangle=\langle L_3\rangle=0,
\end{equation}
where $m'=m+p p_h/2$. Because of this, the operator of the total angular momentum $\textbf{J}=\textbf{L}+\textbf{S}$ has the mean value depending only on the parameters $m, p$, and $p_h$ as
\begin{equation}\label{mtotal}
    \langle J_1\rangle=\hbar m'=\hbar\left(m+\frac{pp_h}{2}\right), \langle J_2\rangle=\langle J_2\rangle=0.
\end{equation}

\begin{figure}
\includegraphics{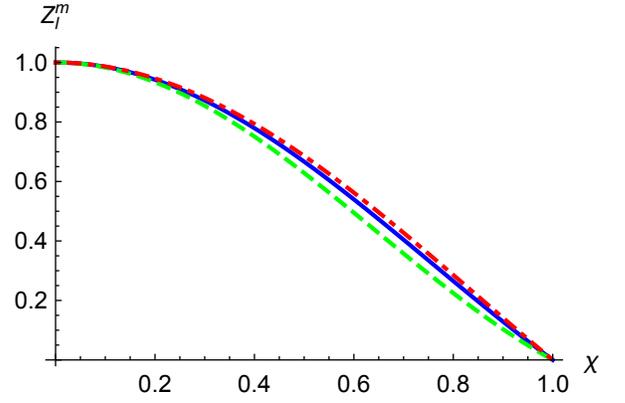}
\caption{\label{fig21pl01p}Plot of $Z_l^m$ against $\chi$ for $l=m=0$ (solid line), $l=1, m=0$ (dashed line), and $l=m=1$ (dash-and-dot line).}
\end{figure}

The operators of Hamilton, velocity, and momentum have the following mean values:
\begin{eqnarray}\label{mHvp}
   \langle H\rangle&=&\frac{m_e c^2}{4\delta q}\left[q_b q_{4b}-q_a q_{4a}+\sinh^{-1}{q_b}-\sinh^{-1}{q_a}\right],\nonumber\\
  \langle v_1\rangle&=&c\frac{q_a+q_b}{q_{4a}+q_{4b}}Z_l^{|m|}(\chi), \langle v_2\rangle=\langle v_3\rangle=0,\nonumber\\
  \langle p_1\rangle&=&m_e c\frac{q_a+q_b}{2}Z_l^{|m|}(\chi), \langle p_2\rangle=\langle p_3\rangle=0,
\end{eqnarray}
where $q_{4a}=\sqrt{1+q_a^2}$ and $q_{4b}=\sqrt{1+q_b^2}$. The dependence of $Z_l^m$ on $\chi$ is shown in Fig.~\ref{fig21pl01p}.

%%%%%%%%%%%%%%%%%%%%%%%%%%%%%%figure21%%%%%%%%%%%%%%%
%%%%%%%%%%%%%%%%%%%%%%%%%%%%%%%%%%%%%%%%%%%%%%%%%%

For the localized solutions $\Psi_{l,p}^m$ illustrated in Figs.~\ref{fig8pla001}--\ref{fig10pla111} and Figs.~\ref{fig12plb111}--\ref{fig14plc001}, $Z_l^{|m|}(1)=0$ and in consequence the mean values of the operators of spin, velocity, and momentum vanish. These mean values are non-zero for the solutions illustrated in Figs.~\ref{fig11pld001} and Figs.~\ref{fig15v1x23ple}--\ref{fig20v3x12ple}, since $Z_0^0(0.5)=2/3$, $Z_1^0(0.5)=22/35$, and $Z_1^1(0.5)=24/35$. The operators of the orbital and total angular momentums have coinciding non-zero mean values $\langle L_1\rangle=\langle J_1\rangle=\hbar$ and $\langle L_1\rangle=\langle J_1\rangle=\hbar(1-p/2)$ with respect to the wave functions $\Psi_{1,p}^1$ depicted in Fig.~\ref{fig10pla111} and Figs.~\ref{fig12plb111}, \ref{fig13plb112}, respectively.

\subsubsection{Wavelet representation}
Since the spinors $v_p$ Eq.~(\ref{rhoN0}) are independent of $q$, whereas the function $\nu$ Eq.~(\ref{nuqchi}) is independent of $q$ and $\varphi$, the wave function $\Psi_{l,p}^m$ Eq.~(\ref{Psilmp}) can be written as
\begin{equation}\label{Psiwavelet}
    \Psi_{l,p}^m=\int_0^\pi d\vartheta \sqrt{\chi\sin\vartheta\sin{\chi\vartheta}}\int_0^{2\pi} d\varphi Y_l^m {\cal W}_p,
\end{equation}
where
\begin{equation}\label{calWp}
    {\cal W}_p=e^{i\Phi_0}\left(
                            \begin{array}{c}
                              f_1 v_p \\
                              p f_2 v_p \\
                            \end{array}
                          \right),
\end{equation}
\begin{equation}\label{f1f2}
    \left(
       \begin{array}{c}
         f_1 \\
         f_2 \\
       \end{array}
     \right)=\frac{1}{\sqrt{2\delta q}}\int_{-\delta q}^{\delta q} \frac{q}{N_0}\left(
                                                                                  \begin{array}{c}
                                                                                    1 \\
                                                                                    \varrho_0 \\
                                                                                  \end{array}
                                                                                \right)e^{i\delta\Phi}d\eta,
\end{equation}
\begin{equation}\label{Phi0q40}
    \Phi_0=2\pi(q_0 R_q -q_{40}X_4),\quad q_{40}=\sqrt{1+q_0^2},
\end{equation}
\begin{eqnarray}\label{dPhidq4}
  \delta\Phi&=&2\pi(\eta R_q -\delta q_4 X_4),\quad \eta=q-q_0,\nonumber\\
  \delta q_4&=&q_4-q_{40}=\frac{\eta(2q_0+\eta)}{q_{40}+q_4}.
\end{eqnarray}
These relations describe the wave function $\Psi_{l,p}^m$ as a superposition of plane wavelets ${\cal W}_p$ with the wave normals
\begin{equation}\label{eqvar}
    \textbf{e}_q=\textbf{e}_1\cos{\chi\vartheta}+\sin{\chi\vartheta}(\textbf{e}_2\cos\varphi+\textbf{e}_3\sin\varphi),
\end{equation}
where $\vartheta\in [0,\pi]$ and $\varphi\in [0,2\pi]$. The wavelet ${\cal W}_p={\cal W}_p(\vartheta,\varphi,R_q,X_4)$ is the wave packet obtained by integrating the corresponding plane harmonic wave function on the quasimomentum $q\in [q_a,q_b]$. As a first approximation for quasimonochromatic beams with $\delta q \ll q_0$, Eq.~(\ref{f1f2}) can be written as
\begin{equation}\label{f1f2ini}
    \left(
      \begin{array}{c}
        f_1 \\
        f_2 \\
      \end{array}
    \right)=\sqrt{2\delta q}\frac{q_0}{N_{00}}\left(
                                                \begin{array}{c}
                                                  1 \\
                                                  \varrho_{00} \\
                                                \end{array}
                                              \right)\frac{\sin D_q}{D_q},
\end{equation}
where
\begin{eqnarray}\label{Dq}
   D_q&=&2\pi {\delta q}\left(R_q -\frac{q_0}{q_{40}}X_4\right),\nonumber\\
  \varrho_{00}&=&\frac{q_0}{q_{40}+1},\quad N_{00}=\sqrt{1+\varrho_{00}^2}.
\end{eqnarray}

\subsection{\label{sec:pack}Wave packets in 2D-ESTC}
In this section, we present the wave packets $\Psi_s$ which can be composed from the basis wave functions $\Psi=\Psi(s,q_n)$ defined by Eq.~(\ref{VWX4}) for $\{L,j,n\}=\{-1,4,1\}$ and by Eq.~(\ref{VWX1}) for $\{L,j,n\}=\{1,4,1\}$ as
\begin{equation}\label{Psis}
    \Psi_s=\nu_s\int_{q_{a,s}}^{q_{b,s}}\Psi(s,q_n)dq_n, \quad \nu_s=\frac{1}{\sqrt{q_{b,s}-q_{a,s}}},
\end{equation}
where $\Psi(s,q_n)$ satisfy the normalization condition Eq.~(\ref{normcond}), and $q_{b,s}-q_{a,s}<\Omega$. Owing to Eqs.~(\ref{V0W0}) and (\ref{abcdforp}), these wave packets satisfy the normalization condition
\begin{equation}\label{Psisnorm}
    {\cal J}_{jn}(\Psi_s^\dag \Psi_s)=1,
\end{equation}
where
\begin{equation}\label{Jjnf}
    {\cal J}_{jn}(f)=\frac{1}{2\pi}\int_0^{2\pi}d\varphi_j\int_{-\infty}^{+\infty}f(\varphi_j,X_n)dX_n.
\end{equation}
The mean value $\langle L \rangle_s$ of a linear operator $L$ with respect to the wave function $\Psi_s$ can be expressed in terms of mean values $\langle L \rangle$ of $L$ with respect to the basis wave functions $\Psi$ as
\begin{equation}\label{Psismean}
    \langle L \rangle_s={\cal J}_{jn}(\Psi_s^\dag L \Psi_s)=\frac{1}{q_{b,s}-q_{a,s}}\int_{q_{a,s}}^{q_{b,s}}\langle L \rangle dq_n.
\end{equation}
To this end, one can use the mean values $\langle L \rangle$ presented in Sec.~\ref{sec:mean}: the velocity $V_1$ Eq.~(\ref{mV1}), the momentum $P_1$ Eq.~(\ref{mP1}), the spin $\langle S_1 \rangle$ Eq.~(\ref{mS1}), and the energy $E$ Eq.~(\ref{mV1}).

In Ref. \cite{ESTCp5}, the superpositions of two basic wave functions $\Psi(-1,q_1)$ and $\Psi(1,q_1)$ describing different spin states and corresponding to (i) the same quasimomentum $q_1$ (unidirectional electron states with the spin precession) and (ii) the two equal-in-magnitude but oppositely directed quasimomenta (bidirectional electron states) are presented. Such electron states can be extended to the wave packets $\Psi_s$ Eq.~(\ref{Psis}) as follows:
\begin{equation}\label{bipacket}
    \Psi_2=\Psi_{-1} e^{i\delta}\cos\alpha+\Psi_1\sin\alpha,
\end{equation}
where $\alpha\in [0,\pi/2]$ and $\delta\in [0,2\pi]$. Let $q_{n-}$ and $q_{n+}$ be integration variables for $\Psi_{-1}$ and $\Psi_{1}$, respectively. If the condition $|q_{n+}-q_{n-}|<\Omega$ is fulfilled for any $q_{n\pm}\in[q_{a,\pm 1},q_{b,\pm 1}]$, then the wave function $\Psi_2$ Eq.~(\ref{bipacket}) satisfies the normalization condition ${\cal J}_{jn}(\Psi_2^\dag \Psi_2)=1$, and the mean value $\langle L \rangle_2$ of a linear operator $L$ with respect to $\Psi_2$ is given by
\begin{equation}\label{Psi2mean}
    \langle L \rangle_2={\cal J}_{jn}(\Psi_2^\dag L \Psi_2)=\langle L \rangle_{-1}\cos^2\alpha +\langle L \rangle_1\sin^2\alpha.
\end{equation}

\subsubsection{2D-ESTC with L=-1}
As an example let us consider first the wave packets $\Psi_{\mp 1}$ composed of $q_{4\mp}$ solutions illustrated in Fig.~\ref{fig6q1E1mp}. In this case, $L=-1, j=4, n=1$, and the mean value of momentum $P_1$ vanishes at $q_1=\pm p_{10}$ for $q_{4\mp}$ solutions, respectively. To obtain a bidirectional electron state $\Psi_2$, we set $q_{a,-1}=-2p_{10}$, $q_{b,-1}=q_{a,1}=0$, $q_{b,1}=2p_{10}$, $\alpha=\pi/4$, and $\delta=0$. The electron states with the wave functions $\Psi_{\mp 1}$ have the mean values of velocity $V_1=\pm 0.000316462$, the spin $\langle S_1 \rangle=\pm \frac{\hbar}{2}\Sigma_{1m}$, and the same values of the momentum $P_1=0$ and the energy $E=1.00319230$. Because of this, the function $\Psi_2$ describes the electron state with the vanishing mean values of the velocity, the momentum, and the spin.

\begin{figure}
\includegraphics{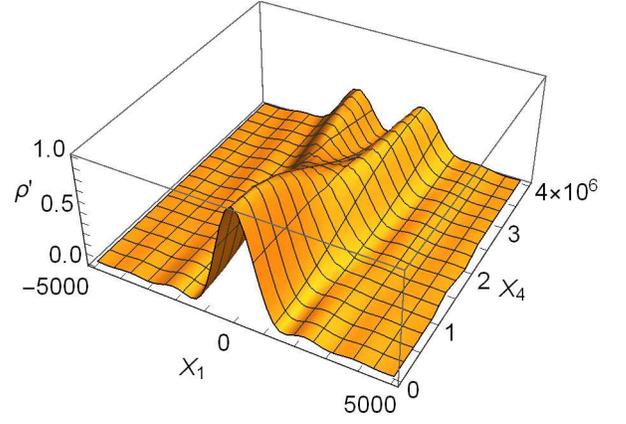}
\caption{\label{fig22rhox1x4}Relative probability density $\rho'=\Psi_2^\dag\Psi_2/\rho_{00}$ as a function of $X_1$ and $X_4$; $\rho_{00}=0.00063392$; the other parameters are the same as described in the caption of Fig.~\ref{fig6q1E1mp}.}
\end{figure}

\begin{figure}
\includegraphics{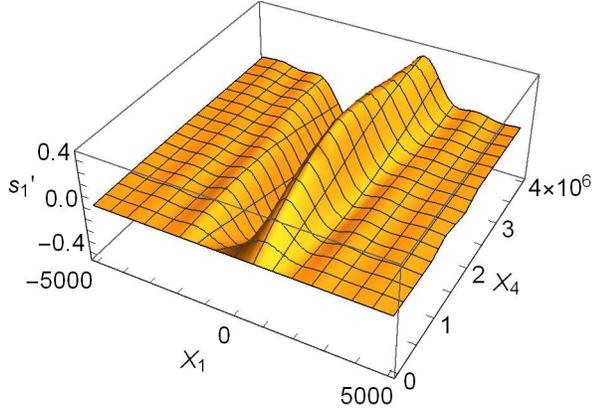}
\caption{\label{fig23s1x1x4}The Hermitian form ${s_1}'=\Psi_2^\dag\Sigma_1\Psi_2/\rho_{00}$ as a function of $X_1$ and $X_4$; the other parameters and notations are the same as described in the captions of Figs.~\ref{fig6q1E1mp} and \ref{fig22rhox1x4}.}
\end{figure}

Figures~\ref{fig22rhox1x4} and \ref{fig23s1x1x4} illustrate the localization at the $X_1$ axis and the evolution with time $X_4$ of the probability density $\rho=\Psi_2^\dag\Psi_2$ and the Hermitian form $\Psi_2^\dag\ S_1\Psi_2=\frac{\hbar}{2}{s_1}'$ of the spin operator $S_1=\frac{\hbar}{2}\Sigma_1$. Whereas in free space an electron has the minimum energy $E=1$ at $q_1=0$, in the 2D-ESTC under consideration it has two different states with the minimum energy $E_{\rm min}$ at $q_1=\pm p_{10}$ and with the mean values of spin opposite in sign (see Sec.~\ref{sec:mean}). In terms of the localized solution $\Psi_2$, this manifests itself with time as the splitting of the central domain with the maximum probability density in two domains with ${s_1}'$ of opposite sign  and negligibly small variations of ${s_1}'$ during the time interval $\Delta X_4=\tau=1/\Omega=10$; see Fig.~\ref{fig23s1x1x4}.

%%%%%%%%%%%%%%%%%%%%%%%%%%%%%figure22%%%%%%%%%%%%%%%
%%%%%%%%%%%%%%%%%%%%%%%%%%%%%%%%%%%%%%%%%%%%%%%%%%

%%%%%%%%%%%%%%%%%%%%%%%%%%%%%figure23%%%%%%%%%%%%%%%
%%%%%%%%%%%%%%%%%%%%%%%%%%%%%%%%%%%%%%%%%%%%%%%%%%

\subsubsection{2D-ESTC, $L=1$}

\begin{figure}
\includegraphics{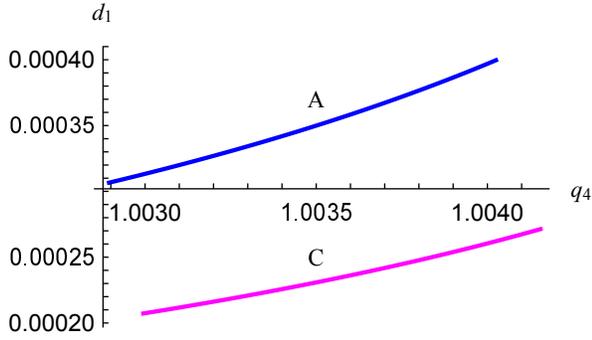}
\caption{\label{fig24q4d1md1p}Function $d_1$ against $q_4$: (A) $d_1=d_{1-}(q_4)$, (C) $d_1=d_{1+}(q_4)$; $L=1$; $g=1$; $\Omega=0.1$; $a_{12}=a_{42}$; $I_A=0.0064$.}
\end{figure}

In the case $L=1, j=4$, and $n=1$, Eqs.~(\ref{dispersion}) and (\ref{d4pm}) are replaced by the relations
\begin{eqnarray}\label{dispersion1}
  q_{1\pm}&=&\sqrt{q_4^2(1+d_{1\pm})^2-1-I_A},\nonumber\\
  q'_{1\pm}&=&\sqrt{(q_4 \pm\Omega)^2(1+d'_{1\pm})^2-1-I_A},
\end{eqnarray}
and
\begin{eqnarray}\label{d1pm}
  d_{1\pm}&=&-1+\frac{1}{q_4}\sqrt{1+I_A+q_{1\pm}^2} ,\nonumber\\
  d'_{1\pm}&=&-1+\frac{1}{q_4 \pm\Omega}\sqrt{1+I_A+{q'}_{1\pm}^2} ,
\end{eqnarray}
respectively. The quasimomentums $q_{1\pm}$ and $q'_{1\pm}$ specify the wave functions $\Psi$ Eq.~(\ref{VWX1}) for a given $q_4$ at $s=\pm 1$. The functions $d_{1\pm}$ and $d'_{1\pm}$ are small ($|d_{1\pm}|\ll 1, |d'_{1\pm}|\ll 1$) and depend only weakly on $q_4$, see Fig.~\ref{fig24q4d1md1p}. They can be found by using the fitness parameter $R$ Eq.~(\ref{RZprime}) in much the same way as $d_{4\pm}$ and $d'_{4\pm}$. The quasimomentums $q_{1\pm}$ and $q'_{1-}$ tend to zero as $q_4$ is reduced. For the 2D-ESTC under consideration, the ground states of the Dirac electron with $q_1=0$ and the vanishing mean values of the operators of velocity and momentum are treated at various values of the frequency $\Omega$ and the intensity $I_A$ in Ref. \cite{ESTCp4}.

%%%%%%%%%%%%%%%%%%%%%%%%%%%%%figure24%%%%%%%%%%%%%%%
%%%%%%%%%%%%%%%%%%%%%%%%%%%%%%%%%%%%%%%%%%%%%%%%%%

%%%%%%%%%%%%%%%%%%%%%%%%%%%%%figure25%%%%%%%%%%%%%%%
\begin{figure}
\includegraphics{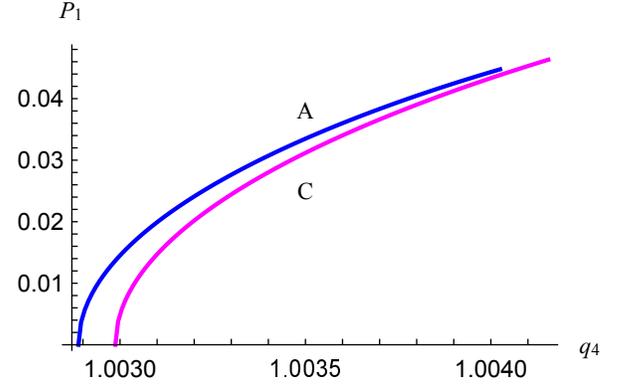}
\caption{\label{fig25q4p1mp1p}Momentum $P_1$ against $q_4$: (A) $s=-1$, (C) $s=1$; the other parameters are the same as described in the caption of Fig.~\ref{fig24q4d1md1p}.}
\end{figure}
%%%%%%%%%%%%%%%%%%%%%%%%%%%%%%%%%%%%%%%%%%%%%%%%%%

Let us now consider the neighborhood of the ground state with $|q_1|<\Omega/2=0.05$. The minimum values $q_{4m-}=1.002887854$ and $q_{4m+}=1.002987379$ of $q_4$ are specified by the condition $q_1=0$ for the $q_{1-}$ and $q_{1+}$ solutions, respectively. Figures \ref{fig24q4d1md1p} and \ref{fig25q4p1mp1p} illustrate the functions $d_1=d_{1\pm}(q_4)$ and the mean value $P_1=P_1(q_4)$ Eq.~(\ref{mP1}) of the momentum operator for these two solutions. There are no solutions $q'_{1-}$ in the $q_4$ domain under consideration, whereas solutions $q'_{1+}$ are specified by Eq.~(\ref{dispersion1}), where the function $d'_{1+}$ can be linearly approximated as
\begin{equation}\label{d1ap}
    d'_{1+}=0.0001631225(q_4-1.0828569),
\end{equation}
with deviations smaller than $10^{-9}$.

The mean values of spin are independent of $q_4$ and take the following values: $\langle S_1 \rangle=-\frac{\hbar}{2}\Sigma_{1m}$ for $q_{1+}$ solution, $\langle S_1 \rangle=\frac{\hbar}{2}\Sigma_{1m}$ for $q_{1-}$ and $q'_{1+}$ solutions, where $\Sigma_{1m}=0.99366079$ coincides with the similar parameter for $q_{4\pm}$ and $q'_{4\pm}$ solutions at $L=-1$. The normalized energy $E$~(\ref{mE}) linearly depends on $q_4$ as
\begin{eqnarray}\label{EL1}
  E&=&q_4 \pm \delta E  \text{ for } q_{1\pm}\text{ solution},\nonumber\\
   &=&q_4 + \Omega -\delta E  \text{ for } q'_{1+}\text{ solution},
\end{eqnarray}
where $\delta E=0.000316960$. The mean values $\langle S_1 \rangle$ and $E$, as well as the functions $d_{1\pm}$ and $d'_{1\pm}$ are independent of the sign of $q_1$. The mean values of the operators of velocity and momentum have the same sign as $q_1$.

Let us now consider the bidirectional superposition $\Psi_2$ Eq.~(\ref{bipacket}) of the wave packets $\Psi_{\pm 1}$ Eq.~(\ref{Psis}) obtained by integrating $\Psi(\pm 1,q_4)$ over the $q_4$ domains illustrated in Figs.~\ref{fig24q4d1md1p} and \ref{fig25q4p1mp1p}. We use the negative branch of the square root in Eq.~(\ref{dispersion1}) for $q_{1-}$, the positive one for $q_{1+}$ and  set $q_{a,-1}=q_{4m-}$, $q_{b,-1}=q_{4m-}+0.00113$, $q_{a,1}=q_{4m+}$, $q_{b,1}=q_{4m+}+0.00116$.

\begin{figure}
\includegraphics{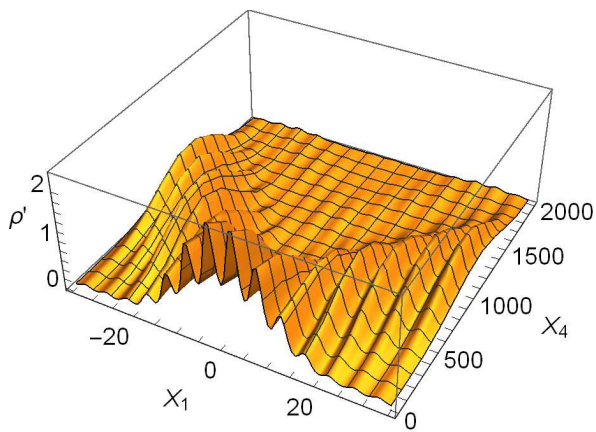}
\caption{\label{fig26rhox1x4}Relative probability density $\rho'=\Psi_2^\dag\Psi_2/\rho_{00}$ as a function of $X_1$ and $X_4$;  $\rho_{00}=0.00063392$; the other parameters are the same as described in the caption of Fig.~\ref{fig24q4d1md1p}.}
\end{figure}

\begin{figure}
\includegraphics{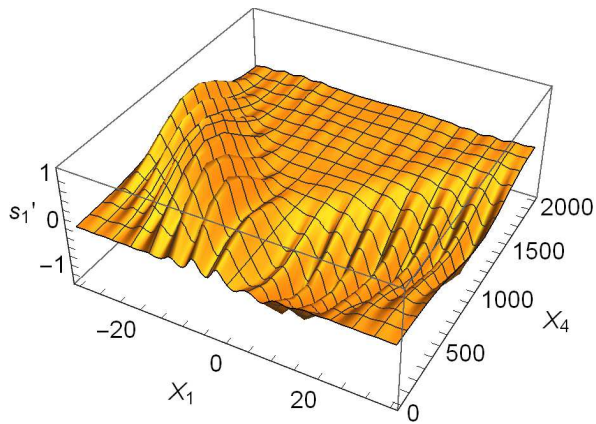}
\caption{\label{fig27s1x1x4}The Hermitian form ${s_1}'=\Psi_2^\dag\Sigma_1\Psi_2/\rho_{00}$ as a function of $X_1$ and $X_4$; the parameters are the same as described in the captions of Figs.~\ref{fig24q4d1md1p} and \ref{fig26rhox1x4}.}
\end{figure}

As for the described above superposition $\Psi_2$ in 2D-ESTC with $L=-1$, the both packets have the same range of $q_1$ magnitudes: $|q_{1\mp}|<0.05$. However, the velocity, momentum, and energy operators now have different magnitudes of mean values $V_{1\mp}$, $P_{1\mp}$, and $E_{\mp}$ with respect to the functions $\Psi_{-1}$ and $\Psi_{1}$:
\begin{eqnarray*}
  V_{1-}&=&-0.030350831,\quad V_{1+}=0.031237500, \\
  P_{1-}&=&-0.030449571,\quad P_{1+}=0.031362632\\
  E_{-} &=&1.0031429,\quad E_{+}=1.0038919.
\end{eqnarray*}
Since the mean values of the spin operator with respect to $\Psi_{-1}$ and $\Psi_{1}$ are equal in magnitude and opposite in direction, the bidirectional state $\Psi_2$ Eq.~(\ref{bipacket}) with the parameters $\alpha=\pi/4$, and $\delta=0$ has the vanishing mean value of spin, but nonvanishing mean values of velocity $V_1=0.00044333427$ and momentum $P_1=0.00045653041$. This localized state has the energy $E=1.0035174$. Figures~\ref{fig26rhox1x4} and \ref{fig27s1x1x4} illustrate the splitting of the central domain with the maximum probability density in two domains with ${s_1}'$ of opposite sign.

%%%%%%%%%%%%%%%%%%%%%%%%%%%%%figure26%%%%%%%%%%%%%%%
%%%%%%%%%%%%%%%%%%%%%%%%%%%%%%%%%%%%%%%%%%%%%%%%%%

%%%%%%%%%%%%%%%%%%%%%%%%%%%%%figure27%%%%%%%%%%%%%%%
%%%%%%%%%%%%%%%%%%%%%%%%%%%%%%%%%%%%%%%%%%%%%%%%%%

\section{Conclusion}
To construct localized solutions of the Dirac equation in the ESTCs or free space, it is necessary to calculate first the basis wave functions $\Psi$ Eq.~(\ref{sol1}) specified by a set of four-dimensional vectors $\textbf{Q},=(\textbf{q},i q_4)$. To attain these ends in the general case of 4D-ESTCs one can use the solutions and techniques presented in Refs.~\cite{bian04,ESTCp1,ESTCp2,ESTCp3,ESTCp4}. It is shown in Sec.~\ref{sec:orthorel} that these wave functions satisfy the orthogonality relations  Eq.~(\ref{orthorel}). In free space, the basis functions reduce to the plane waves $\Psi$ Eq.~(\ref{planew}) which obey the dispersion equation $q_4^2=1+\textbf{q}^2$.

It is shown in Sec.~\ref{sec:2D} that there exist families of wave functions $\Psi$ Eq.~(\ref{VWX4}) and $\Psi$ Eq.~(\ref{VWX1}) in the 2D-ESTCs created by two counterpropagating circularly polarized plane waves, for which the Dirac equation reduces to matrix ordinary differential equations. If these two  electromagnetic waves have the same circular polarization [$L=-1$ in Eq.~(\ref{A1A4})], then Eq.~(\ref{dVWdX4}) defines amplitudes $V=V(X_4)$ and $W=W(X_4)$ of the wave function $\Psi$ Eq.~(\ref{VWX4}). However, if they have left and right circular polarizations ($L=1$), then Eq.~(\ref{dVWdX1}) defines amplitudes $V=V(X_1)$ and $W=W(X_1)$ of the function $\Psi$ Eq.~(\ref{VWX1}). The dispersion equations for both families can be written in the form $q_4^2(1+d)^2=1+q_1^2$, where $|d|\ll 1$. The technique presented in Sec.~\ref{sec:2D} is based on the use of the fitness criterion $R$ Eq.~(\ref{Rpsi}) and Fourier expansions of amplitudes $V$ and $W$. It makes possible to calculate with any prescribed accuracy the basis wave functions $\Psi$ Eq.~(\ref{VWX4}) for a given $q_1$ at $L=-1$ as well as $\Psi$ Eq.~(\ref{VWX1}) for a given $q_4$ at $L=1$. In this article we present integrals of motion and mean values of velocity, momentum, energy, and spin operators with respect to these wave functions.

In Sec.~\ref{sec:local4D}, we extend the general approach to designing and characterizing localized solutions of wave equations~\cite{pre00,pre01,pre02} to the Dirac equation in the ESTC and free space. The presented technique uses the basis wave functions to compose a set of orthonormal beams and various localized states with complex vortex structure of probability currents, defined by a given set of orthonormal complex scalar functions on a two-dimensional manifold. By way of illustration various localized solutions in free space, defined by the spherical harmonics, are presented in Sec.~\ref{sec:free}.

To compose a localized solution of the basis wave functions $\Psi$, one must specify their vectors $\textbf{Q}$, the normalized amplitudes (bispinors in free space and multispinors in the ESTC), magnitudes given by real scalar factors, and initial phases. At a given $\textbf{Q}$, the amplitude subspace is one-dimensional in the 4D-ESTC and the 2D-ESTC treated in this article. Since it is two-dimensional in free space, we defined two linearly independent amplitudes for each given $\textbf{Q}$ and obtained two families of beams $\Psi_{l,p}^m$ with $p=\pm 1$, defined by the spherical harmonics $Y_l^m$. They constitute the ortonormal system satisfying Eq.~(\ref{ortholmp}) and in consequence can be used as a basis in characterizing and designing even more complicated localized solutions. These beams have high probability density only in very small core regions; see Figs.~\ref{fig8pla001}--\ref{fig14plc001}. The beams $\Psi_{l,p}^m$ with $m\neq 0$ have complex vortex structures of probability currents, see Figs.~\ref{fig15v1x23ple}--\ref{fig20v3x12ple}. We also presented the beams localized with respect to all four space-time coordinates, which can be described as flash electron states; see Fig.~\ref{fig14plc001}.

The solutions $\Psi$ Eq.~(\ref{psiq1q4}) are special cases of the function $\Psi$ Eq.~(\ref{sol1}), for which the Fourier expansions of the bispinor amplitude $\Psi_0$ are specified by the subsets ${\cal L}_{-1}=\{(0,0,0,2k),(s,0,0,2k+1);k=0,\pm 1,\pm 2,\ldots\}\subset{\cal L}$ in the 2D-ESTCs with $L=-1$ and ${\cal L}_1=\{(2k,0,0,0),(2k+1,0,0,-s);k=0,\pm 1,\pm 2,\ldots\}\subset{\cal L}$ in the 2D-ESTCs with $L=1$. They are described by the dispersion relations $q_4=q_4(s,q_1)$ and $q_1=q_1(s,q_4)$ with given $q_1$ and $q_4$, respectively. The mean values of velocity, momentum, energy, and spin operators with respect to the both families of basis wave functions and the one-dimensionally localized wave packets are obtained. The similarities and distinctions of these 2D-ESTCs  are illustrated also in terms of the bidirectional electron states.

\begin{acknowledgments}
We thank the anonymous referees for important and useful comments, which were used to revise this article.
\end{acknowledgments}

\bibliography{BorzdovGN_2016_a}

\end{document}